\documentclass{elsartCT}

\usepackage{amssymb}
\usepackage{graphicx}

\begin{document}

\begin{frontmatter}



\title{Exact solutions for the two- and all-terminal reliabilities of a simple ladder network}


\author{Christian Tanguy}

\address{France Telecom R\&D/CORE/MCN/OTT, 38--40 rue du G\'{e}n\'{e}ral Leclerc, 92794 Issy-les-Moulineaux Cedex, France}
\ead{christian.tanguy@orange-ft.com}


\begin{abstract}
The exact calculation of network reliability in a probabilistic context has been a long-standing issue of practical importance, but a difficult one, even for planar graphs, with perfect nodes and with edges of identical reliability $p$. Many approaches (determination of bounds, sums of disjoint products algorithms, Monte Carlo evaluations, studies of the reliability polynomials, etc.) can only provide approximations when the network's size increases.

We consider here a ladder graph of {\em arbitrary} size corresponding to real-life network configurations, and give the exact, analytical solutions for the all- and two-terminal reliabilities. These solutions use transfer matrices, in which {\em individual} reliabilities of edges and nodes are taken into account.
The special case of identical edge and node reliabilities --- $p$ and $\rho$, respectively --- is solved. We show that the zeros of the two-terminal reliability polynomial exhibit structures which differ substantially for seemingly similar networks, and we compare the sensitivity of various edges. We discuss how the present work may be further extended to lead to a catalog of exactly solvable networks in terms of reliability, which could be useful as elementary bricks for a new and improved set of bounds or benchmarks in the general case.



\end{abstract}

\end{frontmatter}

\section{Introduction}
\label{Introduction}

Network reliability has long been a practical issue, and will remain so for years, since networks have now entered an era of Quality of Service (QoS). IP networks, mobile phone networks, transportation networks, electrical power networks, etc., have become ``commodities.'' Connection availability rates of 99.999\% are now an objective for telecommunication network operators, and premium services may only be deployed --- and correspondingly billed --- if the connection reliability is close enough to unity. Reliability is therefore a crucial parameter in the design and analysis of the various kinds of networks we daily use.



Not surprisingly, the study of network reliability has led to a huge body of literature, starting with the work of Moore and Shannon \cite{MooreShannon56}, and including excellent textbooks and surveys \cite{Ball95,Barlow65,Colbourn87,Shier91,Shooman68}. In what follows, we consider a probabilistic approach, in which the network is represented by an undirected graph $G = (V,E)$, where $V$ is a set of nodes (also called vertices) and $E$ is a set of undirected edges (or links), each of which having a probability $p_n$ or $p_e$ to operate correctly. Failures of the different constituents are assumed to occur at random, and to be statistically independent events. Among the different measures of reliability, one often considers the $k$-terminal reliability, namely the probability that a given subset $K$ of $k$ nodes ($K \subset E$) are connected. The most common instances are the all-terminal reliability ${\rm Rel}_A$ ($K \equiv E$) and the two-terminal reliability ${\rm Rel}_2(s \rightarrow t)$, which deals with a particular connection between a source $s$ and a destination $t$. Both of them are affine functions of each $p_n$ and $p_e$.

The sheer number of possible system states, namely $2^{|E|+|V|}$, clearly precludes the use of an ``enumeration of states'' strategy for realistic networks, and shows that the final expression may be extremely cumbersome. Consequently, most studies have considered graphs with perfect nodes ($p_n \equiv 1$) and edges of identical reliability $p$; radio broadcast networks have also been described by networks with perfectly reliable edges but imperfect nodes \cite{AboElFotoh89,Graver05}. It was shown early on --- see for instance the discussion in \cite{Colbourn87,Welsh93} --- that the calculation of $k$-terminal reliability is \#P-hard in the general case, even with the following simplifying and restricting assumptions that (i) the graph is planar (ii) all nodes are perfectly reliable (iii) all edges have the same reliability $p$. All reliabilities are then expressed as a polynomial in $p$, called the reliability polynomial.

The difficulty of the problem has stimulated many approaches: partitioning techniques \cite{Dotson79,Yoo88}, sum of disjoint products \cite{Abraham79,Balan03,Heidtmann89,Rai95,RauzyChatelet03,Soh93}, graph simplifications (series-parallel reductions \cite{MooreShannon56}, delta-wye transformations \cite{Chari96,Egeland91,Gadani81,Rosenthal77,Wang96}, factoring \cite{KevinWood85}), determination of various lower and upper bounds to the reliability polynomial \cite{Ball95,Beichelt89b,BrechtColbourn86,BrownColbourn96,Chen04,Colbourn87,ScottProvan86}, Monte-Carlo simulations \cite{Fishman86,Karger01,Nel90}, genetic \cite{Coit96} and ordered binary decision diagram (OBDD) algorithms \cite{Kuo99,Rauzy03,Yeh02,Yeh02conf}. The reliability polynomial has also been extensively studied \cite{Chari97,Colbourn93,Oxley02}, with the aim of deriving some useful and hopefully general information from the structure of its coefficients \cite{Chari97,Colbourn93} or the location of its zeros in the complex plane \cite{BrownColbourn92}.


A recent breakthrough has been obtained by statistical physicists, who observed that the Tutte polynomial $T(G,x,y)$ of a graph $G$ is actually equivalent to the Potts model partition function of the $q$-state Potts model \cite{Shrock00,Welsh00}, which they were able to calculate for various recursive families of graphs \cite{Chang03}. The all-terminal reliability ${\rm Rel}_A(G,p)$ of such graphs is then deduced from $T(G,1,\frac{1}{1-p})$.
Royle and Sokal \cite{Royle04} proved that the Brown-Colbourn conjecture \cite{BrownColbourn92} on the location of the zeros of the all-terminal reliability polynomial, while valid for series-parallel reducible graphs, does not hold for some families of graphs. While these results are extremely valuable to understand a few properties of graphs and all-terminal reliability polynomials, they still assume that nodes are perfect and that edges have the same reliability.

In recent years, the tremendous growth of Internet traffic has called for a better evaluation of the reliability of connections in optical networks. This, of course, strongly depends on the connection under consideration. Actual failure rates and maintenance data show that a proper evaluation of two-terminal reliabilities must put node and edge equipments on an equal footing, i.e., both edge (fiber links, optical amplifiers) and node (optical cross-connects, routers) failures must be taken into account. The possibility of node failure has been considered in early papers \cite{AboElFotoh89,Evans86,Hansler74}, to quote but a few. Adaptation of algorithms to include imperfect nodes has been --- sometimes controversially --- addressed \cite{Ke97,Netes96,Theologou91,Torrieri94,Yeh02conf}. The two-variable approach for bounds to the reliability polynomial, by Bulka and Dugan \cite{Bulka94} and Chen and He \cite{Chen04}, is also worth mentioning. In order to be realistic, different edge reliabilities should be used too: for instance, the failure rate of optical fiber links is often assumed to increase with their length.

\begin{figure}[thb]
\hskip1cm
\includegraphics[width=0.75\linewidth]{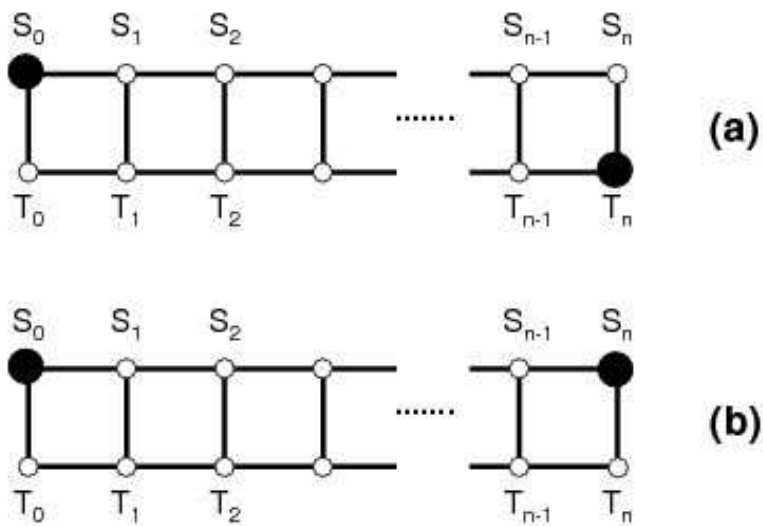}
\vskip0.3cm
\caption{Different source-terminal connections for lattice graphs.}
\label{DeuxEchelles}
\end{figure}

In this work, we give the {\em exact} solution to the two-terminal reliability of a simple ladder network, displayed in Fig.~\ref{DeuxEchelles}, where successive nodes are labelled $S_i$ or $T_j$, and where the larger black dots mark the source $s$ and terminal $t$ we consider in our two-terminal reliability calculations (we choose these special nodes to lie at the graph extremities, since we can always reduce to this case by a series-parallel simplification). This network is a simplified description of a standard (nominal + backup paths) architecture, with additional connections between transit nodes for enabling the so-called ``local protection'' policy, which bypasses faulty intermediate nodes and/or edges. Such an architecture of ``absolutely reliable nodes and unreliable edges,'' with up to 25 edges, was chosen as Example 5 in \cite{Heidtmann89} for a comparison of different ``sum of disjoint products'' minimizing algorithms, or by Rauzy \cite{Rauzy03} as well as Kuo and collaborators \cite{Kuo99,Yeh02,Yeh02conf}. By letting the individual node and edge reliabilities take {\em arbitrary} values, we actually do not add to the complexity of the problem but make the internal structure of the problem more discernible (a similar approach has been fruitful in the context of graph coloring \cite{Biggs72,Biggs01}). Indeed, it is then easier to exploit to the full the recursive nature of the ladder graph while using the delta-wye transformation for graphs with {\em unreliable} nodes \cite{Gadani81}. We show that the two-terminal reliability has a beautiful algebraic structure, as its exact expression is given by a product of $3 \times 3$ transfer matrices
(see eqs.~(\ref{transfermatrixS0Tn}) and (\ref{Rel2S0Tnfinal}) below for the configuration of Fig.~\ref{DeuxEchelles}(a), or  eqs.~(\ref{Rel2S0Snfinal}) and (\ref{transfermatrixS0Sn}) for Fig.~\ref{DeuxEchelles}(b)). Consequently, it can also be determined for an arbitrary size (length) of the network.

\vskip0.6cm
\vskip0.6cm

Our aim is (i) to give a detailed derivation of the final results, so that researchers or engineers involved in reliability studies can readily use an easy-to-implement formula, since worksheet applications are well up to the task (ii) evaluate the implications of these findings, which suggest a new and potentially fruitful approach to reliability in the case of more general graphs or in combinatorial problems such as the enumeration of self-avoiding walks in lattices of restricted width (iii) emphasize anew the importance of algebraic structures of the underlying graphs in the determination of network reliability \cite{Biggs93,Shier91}.


Our paper is organized as follows: In Section~\ref{Triangle-star transformation}, we briefly recall the formulae for the delta-wye transformation for unreliable nodes. In Section~\ref{Main calculation}, we define the notations for the different edge and node reliabilities, used in the detailed derivation of the main results. We then consider in Section~\ref{Identical reliabilities} the case where all edges and nodes have identical reliabilities $p$ and $\rho$, respectively, the size of the network appearing simply as an integer $n$. We give there the analytic solution of the two-terminal reliability ${\rm Rel}_2(p,\rho;n)$ for different configurations, and their asymptotic form when $n \rightarrow \infty$. In Section~\ref{Failures decomposition}, we expand the preceding expressions for $p$ and $\rho$ close to unity, and determine when it is possible to approximate the two-terminal reliability in terms of multiple failures. We also provide in Section~\ref{Generating functions} the associated generating functions, a very useful tool in combinatorics perfectly suited to reliability studies, since they encode all the necessary information in a beautifully simple form. Prompted by the nearly universal character of the Brown-Colbourn conjecture \cite{BrownColbourn92}, we then address the location of zeros of the two-terminal reliability polynomials in Section~\ref{Zeros}, and show that their structures may well be quite distinct even for seemingly similar networks. We then give in Section~\ref{Sensibilite} a short glimpse of the sensitivity issue \cite{Henley91,Rubino91}, namely the influence of a given component to the overall reliability, by comparing the contributions of particular edges (the rungs of the ladder). For the sake of completeness, we derive in Section~\ref{All-terminal reliability} the all-terminal reliability for the ladder networks under consideration, for arbitrary values of edge reliabilities. Finally, we conclude by indicating several directions in which the present results may be further extended, so that, for instance, a catalog of exactly solvable networks --- in terms of reliability --- may be given rapidly \cite{Tanguy06b,Tanguy06c}, which could be useful as elementary bricks for a new and improved set of bounds or benchmarks for alternative methods in the general case.

\section{Triangle-star transformation for unreliable nodes}
\label{Triangle-star transformation}

The triangle-star --- also called delta-star, delta-wye, and $\Delta-{\rm Y}$ --- transformation has been used many times to simplify calculations of network reliability \cite{Chari96,Colbourn87,Egeland91,Gadani81,KevinWood85,Rosenthal77,Wang96}, even though it has mainly been applied in a perfect nodes context, to provide upper and lower bounds to the exact reliability. Here, we fully exploit this transformation in the case of imperfect nodes in order to obtain exact results. Since it plays a crucial part in the derivation, we recall the formulae first derived by Gadani \cite{Gadani81}.

\begin{figure}[htb]
\hskip1cm
\includegraphics[width=0.75\linewidth]{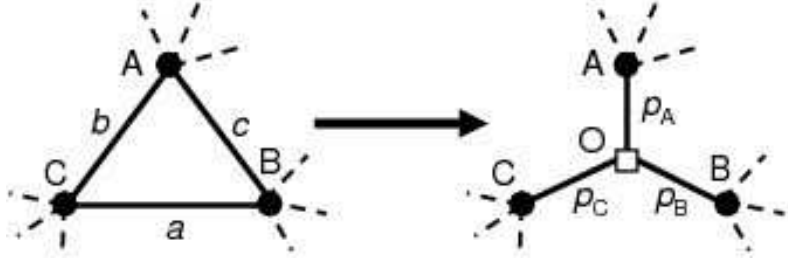}
\vskip0.3cm
\caption{Triangle-star transformation for unreliable nodes. $A$, $B$, and $C$ are the node reliabilities, with $a$, $b$, and $c$ the edge reliabilities of the initial network, and $p_A$, $p_B$, $p_C$, and $O$ those of the transformed network.}
\label{Delta-Y}
\end{figure}

Consider three particular nodes $A$, $B$, and $C$ of a network represented in the left part of Fig.~\ref{Delta-Y} (the reliability of the nodes are given by the same uppercase variables, in order to avoid an unnecessary multiplication of notations). The reliability of the edge connecting $A$ and $B$ is given by (lowercase) $c$, with similar notation for the remaining edges of the triangle. The aim of the triangle-star transformation is to replace the triangle by a star, which is possible by the addition of a new {\em unreliable} vertex $O$ and three new edges connecting $O$ to $A$, $B$ and $C$, with reliabilities $p_A$, $p_B$, and $p_C$, respectively. Both networks are equivalent provided that the following compatibility relations hold \cite{Gadani81}
\begin{eqnarray}
p_A \, O \, p_C & = & b + a \, c \, B - a \, b \, c \, B , \label{Delta-triangle AC}\\
p_A \, O \, p_B & = & c + a \, b \, C - a \, b \, c \, C , \label{Delta-triangle AB}\\
p_B \, O \, p_C & = & a + b \, c \, A - a \, b \, c \, A , \label{Delta-triangle BC}\\
p_A \, O \, p_B \, p_C & = & a \, b + b \, c + a \, c - 2 \, a \, b \, c . \label{Delta-triangle ABC}
\end{eqnarray}

\vskip0.6cm
\vskip0.6cm

Note that the first three equalities correspond to the probability that the two nodes under consideration are connected, while the last gives the probability that all three nodes are connected. A word of caution --- given by Gadani --- is worth mentioning in the case of {\em successive} triangle-star transformations: the triangles should have no common edge or node.

\section{Derivation of the main results}
\label{Main calculation}

\begin{figure}[htb]
\hskip2cm
\includegraphics[width=0.75\linewidth]{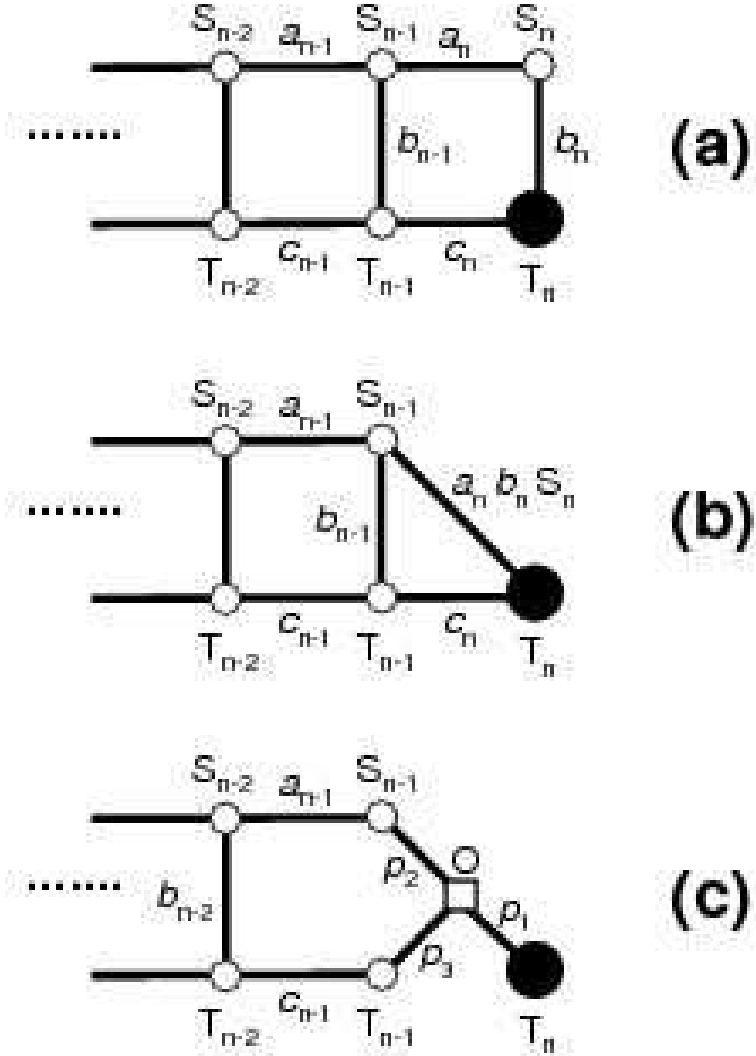}
\vskip0.3cm
\caption{Two-step decomposition of the ladder network from $(a)$ to $(c)$, using the triangle-star transformation.}
\label{Simplification Echelle S0-Tn}
\end{figure}

We first name the different edge and node reliabilities of the ladder diagrams represented in Fig.~\ref{DeuxEchelles}(a), and detail how we can use the triangle-star transformation of the preceding section. Assuming that $S_0$ is always the source node, and using node reliabilities $S_i$ or $T_j$, we  note $a_n$ the reliability of the edge $(S_{n-1},S_{n})$, $b_n$ for $(S_{n},T_{n})$, and finally $c_n$ for $(T_{n-1},T_{n})$, as shown in Fig.~\ref{Simplification Echelle S0-Tn}(a).

\vskip0.6cm
\vskip0.6cm

\subsection{Calculation of ${\rm Rel}_2(S_0 \rightarrow T_n)$}
\label{Rel2S0Tn}

The two-terminal reliability can be written
\begin{equation}
{\rm Rel}_2(S_0 \rightarrow T_n) = T_n \, {{\rm F}_n}(b_0, a_1, b_1, c_1, \cdots , a_n, b_n, c_n) .
\label{Rel2avecFn}
\end{equation}
because the end-node reliabilities obviously factor themselves out of the expression. The transformation from Fig.~\ref{Simplification Echelle S0-Tn}(a) to Fig.~\ref{Simplification Echelle S0-Tn}(b) is a mere series simplification, where a single edge of reliability $a_n \, b_n \, S_n$ is now connecting $S_{n-1}$ to $T_n$. We can then apply the triangle-star transformation to the triangle constituted by $\{S_{n-1},T_{n-1},T_{n}\}$, and define new edges $p_1$, $p_2$, $p_3$, and a node $O$ as represented in Fig.~\ref{Simplification Echelle S0-Tn}(c). From eqs.~(\ref{Delta-triangle AC})--(\ref{Delta-triangle ABC}), we have
\begin{eqnarray}
p_1 \, O \, p_2 & = & a_n \, b_n \, S_n + b_{n-1} \, c_n \, T_{n-1} - a_n \, b_{n-1} \, b_n \, c_n \, S_n \, T_{n-1} \label{p1Op2} , \\
p_1 \, O \, p_3 & = & c_n + b_{n-1} \, a_n \, b_n \, S_{n-1} \, S_{n} - a_n \, b_{n-1} \, b_n \, c_n \, S_{n-1} \, S_{n} , \label{p1Op3}\\
p_2 \, O \, p_3 & = & b_{n-1}  + a_n \, b_n \, c_n \, S_n \, T_n - a_n \, b_{n-1} \, b_n \, c_n \, S_n \, T_n  , \label{p2Op3}\\
p_1 \, O \, p_2 \, p_3 & = & a_n \, b_n \, c_n \, S_n + a_n \, b_{n-1} \, b_n \, S_n +  b_{n-1} \, c_n - 2 \, a_n \, b_{n-1} \, b_n \, c_n \, S_n . \label{p1Op2p3}
\end{eqnarray}
It is then obvious that the two-terminal reliability of the new network displayed in Fig.~\ref{Simplification Echelle S0-Tn}(c) must be {\em formally} similar to the one we started from. The product $T_n \, p_1 \, O$ (a dangling link) is easily factored, and we are left with the calculation of ${\rm Rel}_2(S_0 \rightarrow O)$. This is nothing but the calculation of ${\rm Rel}_2(S_0 \rightarrow T_{n-1})$, where $O$ has replaced $T_{n-1}$ and where adjustments are due: $b_{n-1}$ and $c_{n-1}$ must be replaced by $p_2$ and $c_{n-1} p_3 \, T_{n-1}$, respectively. In short, we have
\begin{equation}
{{\rm F}_{n}} = p_1 \, O \, {{\rm F}_{n-1}}(b_{n-1} \rightarrow p_2, c_{n-1} \rightarrow c_{n-1} p_3 \, T_{n-1}) .
\label{Rel2avecFn-1}
\end{equation}

This relation between ${\rm F}_{n}$ and ${\rm F}_{n-1}$, in which the recursive nature of the ladder graph is now fully apparent, can be further simplified. Indeed, assuming for ${\rm F}_{n}$ the simple form
\begin{equation}
{\rm F}_{n} = \beta_n \, b_n \, S_n + \gamma_n \, c_n + \delta_n \, b_n \, c_n \, S_n
\label{Fn}
\end{equation}
in eqs.~(\ref{Rel2avecFn}) and (\ref{Rel2avecFn-1}), and replacing all occurrences of $p_1$, $p_2$, $p_3$, and $O$, we easily find
\begin{equation}
\left(
\begin{array}{c}
\beta_n \\
\gamma_n \\
\delta_n
\end{array}
\right)
=
M_n
\; \cdot \;
\left(
\begin{array}{c}
\beta_{n-1} \\
\gamma_{n-1} \\
\delta_{n-1}
\end{array}
\right) ,
\end{equation}
with the transfer matrix $M_n$ given by
\begin{equation}
M_n
=
\left(
\begin{array}{c|c|c}
{a_{n}} \, {S_{n-1}} & {a_{n}} \, {b_{n-1}} \, {c_{n-1}} \, {S_{n-1}} \, {T_{n-1}} & {a_{n}} \, {b_{n-1}} \,{c_{n-1}} \, {S_{n-1}} \, {T_{n-1}} \\ \hline
{b_{n-1}}\, {S_{n-1}} \, {T_{n-1}}  &   {c_{n-1}} \, {T_{n-1}} &  {b_{n-1}} \, {c_{n-1}} \, {S_{n-1}} \, {T_{n-1}} \\ \hline
- {a_{n}} \, {b_{n-1}} \, {S_{n-1}} \, {T_{n-1}} & -  {a_{n}} \, {b_{n-1}} \,{c_{n-1}}\, {S_{n-1}} \, {T_{n-1}} & {a_{n}} \, (1 - 2 \, {b_{n-1}}) \,{c_{n-1}}\, {S_{n-1}} \, {T_{n-1}}
\label{transfermatrixS0Tn}
\end{array}
\right) .
\end{equation}
We now need to calculate ${\rm F}_{1}$. This is straightforward, since we have a simple series-parallel graph:
\begin{equation}
{\rm F}_{1} = S_0 \, (a_1 \, b_1 \, S_1 + b_0 \, c_1 \, T_0 - b_0 \, a_1 \, b_1 \, c_1 \, S_1 \, T_0) ,
\end{equation}
from which we deduce $\beta_1 = a_1 \, S_0$, $\gamma_1 = b_0 \, S_0 \, T_0$, and $\delta_1 = - b_0 \, a_1 \, S_0 \, T_0$, or even more simply
\begin{equation}
\left(
\begin{array}{c}
\beta_1 \\
\gamma_1 \\
\delta_1
\end{array}
\right)
=
M_1
\; \cdot \;
\left(
\begin{array}{c}
1 \\
0 \\
0
\end{array}
\right) .
\label{debut S0->T1}
\end{equation}
Note that eq.~(\ref{debut S0->T1}) is correct, even if $c_0$ is undefined; $c_0$ may be set to 0 for the sake of simplicity. We finally obtain the beautifully simple and compact result for the two-terminal reliability between nodes $S_0$ and $T_n$:
\begin{equation}
{\rm Rel}_2(S_0 \rightarrow T_n; n \geq 1) = T_n \, \cdot \; (b_n \, S_n, c_n, b_n \, c_n \, S_n) \; \cdot \;  M_n \; M_{n-1} \; \cdots M_1 \; \cdot \;
\left(
\begin{array}{c}
1 \\
0 \\
0
\end{array}
\right) ,
\label{Rel2S0Tnfinal}
\end{equation}
with $M_n$ given by eq.~(\ref{transfermatrixS0Tn}). We can also write $T_n \, \cdot \; (b_n \, S_n, c_n, b_n \, c_n \, S_n) = (0,1,0) \cdot M_{n+1}$. Note that ${\rm Rel}_2(S_0 \rightarrow T_0) = b_0 \, S_0 \, T_0$ may be obtained by replacing $M_n \cdots M_1$, which is undefined for $n=0$, by the $3 \times 3$ identity  matrix --- or no matrix at all. It is also worth remarking that each edge or node reliability --- which may be arbitrary --- appears once and once only in the different matrices or vectors, respecting the general property that ${\rm Rel}_2(S_0 \rightarrow T_n)$ is an affine function of each component reliability. We have independently checked the above expressions with a sum of disjoint products method for the first values of $n$. Numerically speaking, the formula given in eq.~(\ref{Rel2S0Tnfinal}) is very easy to implement, even in a worksheet application where multiplication of $3 \times 3$ matrices is routinely performed; it also easily applies for ladders of arbitrary size, which is also quite important. This is the final answer to the two-terminal reliability of Example 5 of \cite{Heidtmann89}, which only gave the number of disjoint products for up to 25 unreliable edges, and of benchmark networks \#29--\#30 of \cite{Kuo99,Rauzy03,Yeh02,Yeh02conf}, where ${\rm Rel}_2(S_0 \rightarrow T_n)$ was calculated for $n=19$ and $n=99$ using OBDD algorithms.


\subsection{Calculation of ${\rm Rel}_2(S_0 \rightarrow S_n)$}
\label{Rel2S0Sn}

Let us now turn to the calculation of the two-terminal reliability between two nodes located on the same side of the ladder, as $S_0$ and $S_n$ in Fig.~\ref{DeuxEchelles}(b). The recipe is the same as above: the edges $b_n$ and $c_n$ are merged with the node $T_n$ to give a new equivalent edge between $T_{n-1}$ and $S_n$, of reliability $b_n \, c_n \, T_n$. Here again, we can use the triangle-star transformation to derive new recursion relations, and transfer matrices appear. The calculations being straightforward and similar to those in the preceding subsection, we give only the final result
\begin{equation}
{\rm Rel}_2(S_0 \rightarrow S_n; n \geq 1) = S_n \, \cdot \; (a_n, b_n \, T_n, a_n \, b_n \, T_n) \; \cdot \;  {\widetilde M}_n \; {\widetilde M}_{n-1} \; \cdots {\widetilde M}_1 \; \cdot \;
\left(
\begin{array}{c}
1 \\
0 \\
0
\end{array}
\right) ,
\label{Rel2S0Snfinal}
\end{equation}
with
\begin{equation}
{\widetilde M}_n
=
\left(
\begin{array}{c|c|c}
{a_{n-1}} \, {S_{n-1}} & {b_{n-1}} \, {S_{n-1}} \, {T_{n-1}} & {a_{n-1}} \, {b_{n-1}} {S_{n-1}} \, {T_{n-1}} \\ \hline
{a_{n-1}} \, {b_{n-1}} \, c_n \, {S_{n-1}} \, {T_{n-1}} &  {c_{n}} \, {T_{n-1}} &  {a_{n-1}} \, {b_{n-1}} \, c_n \, {S_{n-1}} \, {T_{n-1}} \\ \hline
- {a_{n-1}} \, {b_{n-1}} \, c_n  \, {S_{n-1}} \, {T_{n-1}} & - {b_{n-1}} \, c_n  \, {S_{n-1}} \, {T_{n-1}} & {a_{n-1}} \, (1 - 2 \, {b_{n-1}}) \,{c_{n}}\, {S_{n-1}} \, {T_{n-1}}
\end{array}
\right)
\label{transfermatrixS0Sn}
\end{equation}
{\em and the additional convention} $a_0 = 1$.  Since $S_n \, \cdot \; (a_n, b_n \, T_n, a_n \, b_n \, T_n) = (1,0,0) \cdot {\widetilde M}_{n+1}$, eq.~(\ref{Rel2S0Snfinal}) also applies in the case $n=0$, which gives ${\rm Rel}_2(S_0 \rightarrow S_0) = S_0$, as it should.

\section{Identical reliabilities $p$ and $\rho$}
\label{Identical reliabilities}

While eqs.~(\ref{Rel2S0Tnfinal}) and (\ref{Rel2S0Snfinal}) give the two-terminal reliability in the general case, we consider here the special case where edges and nodes have reliabilities $p$ and $\rho$, respectively. Our exact results may prove useful because identical edge reliabilities are usually taken for granted, and combinatorial aspects play an important part in the corresponding calculations \cite{Colbourn87}. We show in this section that the two-terminal reliability of the ladder graph has a simple analytic expression for arbitrary $n$, which we derive from the above expansions.

\subsection{Calculation of ${\rm Rel}_2^{(S_0 \rightarrow T_n)}(p,\rho)$}

Taking $a_i = b_i = c_i \equiv p$, and $S_i = T_i \equiv \rho$, we get
\begin{equation}
{\rm Rel}_2^{(S_0 \rightarrow T_n)}(p,\rho) = (p \, \rho^2, p \, \rho, p^2 \, \rho^2) \; \cdot \;
\left(
\begin{array}{c|c|c}
p \, \rho & p^3 \, \rho^2 & p^3 \, \rho^2 \\ \hline
p \, \rho^2 & p \, \rho & p^2 \, \rho^2 \\ \hline
- p^2 \, \rho^2 & - p^3 \, \rho^2 & p^2 \, (1 - 2 \, p) \, \rho^2
\end{array}
\right)^n
 \; \cdot \;
\left(
\begin{array}{c}
1 \\
0 \\
0
\end{array}
\right) .
\label{Rel2S0Tnprhomatrice}
\end{equation}
We could of course replace $(p \, \rho^2, p \, \rho, p^2 \, \rho^2)$ by $(0,1,0)$, provided that the matrix exponent is $n+1$ instead of $n$. The next step is to find the eigenvalues and eigenvectors of the matrix appearing in eq.~(\ref{Rel2S0Tnprhomatrice}), so that the final result may be further simplified. Obviously, a $p^n \, \rho^n$ factor can be factored out of the matrix contribution. We are left with the determination of the eigenvalues and possibly eigenvectors of
\begin{equation}
M_T = \left(
\begin{array}{c|c|c}
1 & p^2 \, \rho & p^2 \, \rho \\ \hline
\rho & 1 & p \, \rho \\ \hline
- p \, \rho & - p^2 \, \rho & p \, (1 - 2 \, p) \, \rho
\end{array}
\right) .
\label{matriceRelS0TnpourPetRho}
\end{equation}
From the first two rows of $M_T$, $1 - p \, \rho$ is obviously an eigenvalue of the matrix. By using standard linear algebra, the characteristic polynomial of $M_T$ can be factored, the three eigenvalues $x_0$ and $x_\pm$ of $M_T$ being simply
\begin{eqnarray}
x_0 & = & 1 - p \, \rho , \label{RacinespourPetRhoX0}\\
x_{\pm} & = & \frac{1 + 2 \, p \,  (1 - p) \, \rho \pm \sqrt{1 + 4 \, p^2 \, \rho - 8 \, p^3 \, \rho^2 +4 \, p^4 \, \rho^2}}{2} .
\label{RacinespourPetRhoXpm}
\end{eqnarray}
While the determination of the eigenvectors may be used to finish the calculation, this is by no means necessary. Indeed, $M_T$ being diagonalizable --- the three roots are distinct --- ${\rm Rel}_2^{(S_0 \rightarrow T_n)}(p,\rho)$ must be of the form
\begin{equation}
{\rm Rel}_2^{(S_0 \rightarrow T_n)}(p,\rho) = p^n \, \rho^n \; \left( \alpha_0 \, x_0^n + \alpha_+ \, x_+^n + \alpha_- \, x_-^n  \right).
\label{FormeRelS0TnpourPetRho}
\end{equation}
From  the first values of ${\rm Rel}_2^{(S_0 \rightarrow T_n)}(p,\rho)$ for $n = 0, 1, 2$
\begin{eqnarray}
{\rm Rel}_2^{(S_0 \rightarrow T_0)}(p,\rho) & = &  p \, \rho^2 , \label{RelS0T0pourPetRho}\\
{\rm Rel}_2^{(S_0 \rightarrow T_1)}(p,\rho) & = &  p^2 \, \rho^3 \, \left( 2 - p^2 \, \rho \right) , \label{RelS0T1pourPetRho}\\
{\rm Rel}_2^{(S_0 \rightarrow T_2)}(p,\rho) & = &  p^3 \, \rho^4 \, \left( 3 - 2 \, p^2 \, \rho + p^2 \, \rho^2 \, (1 - p) \, (1 - 2 \, p) \right) ,
\label{RelS0T2pourPetRho}
\end{eqnarray}
which have been independently checked by a sum of disjoint products procedure, and the inversion of a $3 \times 3$ Vandermonde matrix based on $x_0$ and $x_\pm$, we obtain $\alpha_0$ and $\alpha_{\pm}$. Finally
\begin{eqnarray}
{\rm Rel}_2^{(S_0 \rightarrow T_n)}(p,\rho) & = & \frac{p^n \, \rho^{n+1}}{2} \, \left[- (1-p \, \rho)^{n+1} + (1+p \, \rho) \, \frac{x_+^{n+1} - x_-^{n+1}}{x_+ - x_-} \right. \nonumber \\
& & \hskip2cm \left. - p \, \rho \,  (1- 2 \, p + p \, \rho) \, \frac{x_+^{n}
- x_-^{n}}{x_+ - x_-}\right] .
\label{expansionRelS0TnpourPetRho}
\end{eqnarray}
Setting $p=0.9$ and $\rho = 0.9 {\rm \ or \ } 1$ for $n = 19 {\rm \ or \ } 99$ confirms the numerical results found for the $2 \times 20$ and $2 \times 100$ ladder networks \cite{Kuo99,Rauzy03,Yeh02,Yeh02conf}.

We discuss the implications of eq.~(\ref{expansionRelS0TnpourPetRho}) in section \ref{Discussion des resultats p et rho} below. We will see in Section~\ref{Generating functions} that the simplicity of the result stems from our choosing a unique reliability $p$ for all edges. Before that, we turn to other two-terminal configurations.

\subsection{Calculation of ${\rm Rel}_2^{(S_0 \rightarrow S_n)}(p,\rho)$}

In the $S_0 \rightarrow S_n$ configuration, eqs.~(\ref{Rel2S0Snfinal}) and (\ref{transfermatrixS0Sn}) lead to
\begin{equation}
{\rm Rel}_2^{(S_0 \rightarrow S_n)}(p,\rho) = \rho \, (p, p \, \rho, p^2 \, \rho) \; \cdot \;
\left(
\begin{array}{c|c|c}
p \, \rho & p \, \rho^2 & p^2 \, \rho^2 \\ \hline
p^3 \, \rho^2 & p \, \rho & p^3 \, \rho^2 \\ \hline
- p^3 \, \rho^2 & - p^2 \, \rho^2 & p^2 \, (1 - 2 \, p) \, \rho^2
\end{array}
\right)^n
 \; \cdot \;
\left(
\begin{array}{c}
1/p \\
0 \\
0
\end{array}
\right) .
\label{Rel2S0Snprhomatrice}
\end{equation}
Note that while $a_0 \equiv 1$ is the only $a_i$ which should not be set to $p$, we can compensate this missing $p$ term in ${\widetilde M}_1$ by introducing a $1/p$ term in the rightmost vector in eq.~(\ref{Rel2S0Snfinal}). The procedure is the same as above: a $(p \, \rho)^n$ prefactor can safely be extracted from the matrix on the right-hand side of eq.~(\ref{Rel2S0Snprhomatrice}), so that we have to find the eigenvalues of another matrix, namely $M_S$
\begin{equation}
M_S = \left(
\begin{array}{c|c|c}
1 & \rho & p \, \rho \\ \hline
p^2 \, \rho & 1 & p^2 \, \rho \\ \hline
- p^2 \, \rho & - p \, \rho & p \, (1 - 2 \, p) \, \rho
\end{array}
\right) .
\label{matriceRelS0SnpourPetRho}
\end{equation}
Unsurprisingly, the eigenvalues of $M_S$ are again given by $x_0$ and $x_\pm$.
Finding the prefactors associated with $x_0^n$ and $x_\pm^n$ from the first three values of ${\rm Rel}_2^{(S_0 \rightarrow S_n)}(p,\rho)$ is straightforward, and the final expression is
\begin{eqnarray}
{\rm Rel}_2^{(S_0 \rightarrow S_n)}(p,\rho) & = & \frac{p^n \, \rho^{n+1}}{2} \, \left[+ (1-p \, \rho)^{n+1} + (1+p \, \rho) \, \frac{x_+^{n+1} - x_-^{n+1}}{x_+ - x_-} \right. \nonumber \\
& & \hskip2cm \left. - p \, \rho \,  (1- 2 \, p + p \, \rho) \, \frac{x_+^{n}
- x_-^{n}}{x_+ - x_-}\right] .
\label{expansionRelS0SnpourPetRho}
\end{eqnarray}
It is worth noting that the expressions given in eqs.~(\ref{expansionRelS0TnpourPetRho}) and (\ref{expansionRelS0SnpourPetRho}) only differ by a `$\pm$' sign.

\subsection{Calculation of ${\rm Rel}_2^{(S_0 \rightarrow U_n)}(p,\rho)$}

\begin{figure}[thb]
\hskip1cm
\includegraphics[width=0.75\linewidth]{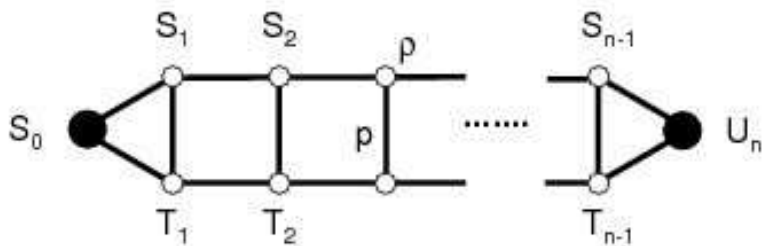}
\vskip0.3cm
\caption{Symmetrical ladder network.}
\label{Echelle Symetrique S0-Un}
\end{figure}

Let us introduce another configuration --- the symmetrical ladder network of Fig.~\ref{Echelle Symetrique S0-Un} --- which looks quite similar to the two previous ones because (i) it has been considered in the literature as a way to describe a ``gamma multistage interconnection network'' \cite{Rai95}, an architecture demonstrating the potential usefulness of recursive algorithms by H\"{a}nsler \cite{Hansler75} or Fratta and Montanari \cite{Fratta78} (ii) it will shed some light on what can be expected from the study of the two-terminal reliability polynomial, as regards the location of its zeros.

\vskip0.6cm
\vskip0.6cm

We do not give the exact expression of ${\rm Rel}_2^{(S_0 \rightarrow U_n)}$ for arbitrary edge and node reliabilities, since it can be derived from eqs.~(\ref{transfermatrixS0Tn}) and (\ref{Rel2S0Tnfinal}) by merely taking $T_0 = S_n = b_0 = b_n = 1$. We limit ourselves to the $(p,\rho)$ configuration and find, after a treatment similar to that of the previous sections,
\begin{eqnarray}
{\rm Rel}_2^{(S_0 \rightarrow U_n)}(p,\rho) & = & \frac{p^n \, \rho^{n+1}}{x_+ - x_-} \, \left[(2 - p) \,
(x_+^{n} - x_-^{n}) \right. \nonumber \\
& & \hskip2cm \left. + p \, (1- 2 \, \rho + p \, \rho) \, (x_+^{n-1} - x_-^{n-1}) \right] .
\label{expansionRelS0UnpourPetRho}
\end{eqnarray}
Note that $x_0$ is absent here: only two eigenvalues appear, even though the network architecture differs from the previous ones by two edges only.

\subsection{Variation of the eigenvalues with $p$ and $\rho$}
\label{Discussion des resultats p et rho}

\begin{figure}[thb]
\hskip2cm
\includegraphics[width=0.75\linewidth]{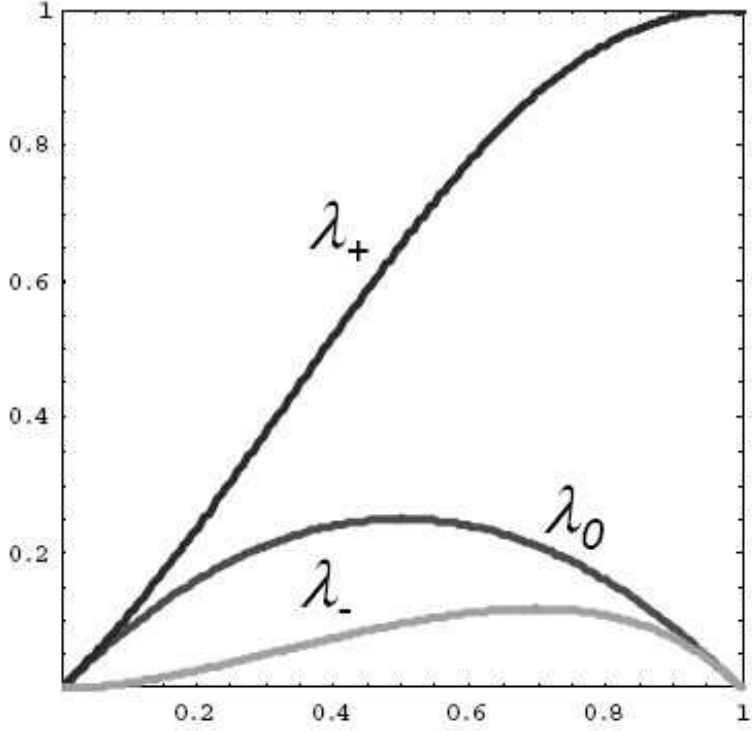}
\vskip0.3cm
\caption{Dependence of $\lambda_0$, $\lambda_+$, and $\lambda_-$ on $p$, when $\rho = 1$.}
\label{Variation des lambdas}
\end{figure}

From eqs.~(\ref{expansionRelS0TnpourPetRho}), (\ref{expansionRelS0SnpourPetRho}), and (\ref{expansionRelS0UnpourPetRho}), it  appears that the two-terminal reliabilities for the ladders are given by sums of terms such as $\lambda_0^n$, $\lambda_+^n$, and $\lambda_-^n$, with
\begin{eqnarray}
\lambda_0 & = & p \, \rho \, (1 - p \, \rho) , \label{definition de lambda 0}\\
\lambda_{\pm} & = & \frac{1}{2} \, p \, \rho \, \bigg( 1 + 2 \, p \, (1 - p) \, \rho \pm \sqrt{1 + 4 \, p^2 \, \rho - 8 \, p^3 \, \rho^2 +4 \, p^4 \, \rho^2} \bigg) ,
\label{definition de lambda +/-}
\end{eqnarray}
where $n$ appears only in the exponents. A natural first step is to assess these quantities as $p$ and $\rho$ vary, in order to assess the relative contributions of the different terms. For the sake of simplicity, we have displayed on Fig.~\ref{Variation des lambdas} the variation of $\lambda_0$, $\lambda_+$, and $\lambda_-$ with $p$, when $\rho = 1$. All three quantities are positive and vanish when $p$ goes to zero, as expected. However, only $\lambda_+$ tends to 1 when $p \rightarrow 1$ (remember that for $p \rightarrow 1$, ${\rm Rel}_2 \rightarrow 1$, implying that one of the eigenvalues tends to 1). In this particular case, we have $\lambda_+ > \lambda_0 > \lambda_-$ in the entire range $0 < p < 1$.

\vskip0.6cm
\vskip0.6cm

An important issue is the behavior of ${\rm Rel}_2^{(S_0 \rightarrow T_n, S_n, U_n)}(p,\rho)$ as the size of the network --- or $n$ --- increases. We see on Fig.~\ref{Variation des lambdas} that for $p = 0.9$, $\lambda_+$ is about ten times larger than $\lambda_0$ and $\lambda_-$. Consequently, when $n$ is sufficiently large --- this may be valid for $n \sim 10$ already  --- we have to a very good approximation
\begin{equation}
{\rm Rel}_2^{(S_0 \rightarrow T_n, S_n, U_n)}(p,\rho) \approx \alpha_+ \, \lambda_+^n ,
\label{limite asymptotique}
\end{equation}
which is observed in the numerical evaluations of Kuo {\em et al.} and Rauzy \cite{Kuo99,Rauzy03,Yeh02,Yeh02conf}. Basically, adding one ``cell'' to the ladder network will merely result in another $\lambda_+$ factor in the two-terminal reliability, so that $\lambda_+$ represents a scaling of the reliability; it is as if we had $n$ equipments of reliability $\lambda_+$ in series. When $\rho \neq 1$, this general behavior is the same, even if $\lambda_-$ is negative for $\frac{1}{2-\rho} \leq p \leq 1$, $\lambda_+$ still prevails, and the power-law behavior of eq.~(\ref{limite asymptotique}) holds.

\section{Failures decomposition}
\label{Failures decomposition}

In some instances, when $p$ and $\rho$ are assumed to be very close to 1, calculations often involve the unreliability of the connection $U_2 = 1 - {\rm Rel}_2$. Let $q = 1-p$ and $\eta = 1 - \rho$. The network unreliability is then expanded in $q$ and $\eta$, even though one should be careful about such a practice, when $q$ and $\eta$ differ by orders of magnitude. For instance, keeping only three terms in $U_2$, we find from eq.~(\ref{expansionRelS0TnpourPetRho}) that the following formula is valid for $n$ greater than 3 (but not too large, of course)
\begin{eqnarray}
U_2(S_0 \rightarrow T_n) & = & 2 \, \eta + (3 \, n - 4) \, \eta^2 - (12 \, n - 16) \, \eta^3 + 4 \, n \, \eta \, q \nonumber \\
& & - (20 \, n - 6) \, \eta^2 \, q+ (n+2) \, q^2 - (4 \, n + 14) \, \eta \, q^2 \nonumber\\
& & + 2 \, (n-1) \, q^3 + \cdots
\label{U trois pannes}
\end{eqnarray}
We must keep in mind, however, that $q$ and $\eta$ must be sufficiently small for such an expansion to make sense. Limiting ourselves to $\eta= 0$ (the perfect nodes configuration), we get
\begin{equation}
U_2(S_0 \rightarrow T_n) = (n+2) \, q^2 + 2 \, (n-1) \, q^3 - \frac{n^2 + 7 \, n +10}{2} \, q^4 + \cdots
\label{U 4 pannes pour Sn et Tn}
\end{equation}
with the same result for $U_2(S_0 \rightarrow S_n)$, whereas
\begin{equation}
U_2(S_0 \rightarrow U_n) = n \, q^2 + 2 \, (n-1) \, q^3 - \frac{n \, (n+3)}{2} \, q^4 + \cdots
\label{U 4 pannes pour Sn et Tn}
\end{equation}
The ``small parameter'' in the expansion is therefore of the order of $n \, q$. A rule of thumb is to ensure this quantity is small indeed when considering expansions such as eq.~(\ref{U trois pannes}).

\section{Generating functions}
\label{Generating functions}

There is yet another way to express all the abovementioned results is a simple and compact form, namely that of the so-called ``ordinary generating functions'', which are a fundamental tool in combinatorics \cite{Stanley97} and extensively used in \cite{Chang03}. The matrix transfer formalism and results obtained in the preceding sections clearly indicate the existence of recursion relations between successive reliability polynomials (see \cite{Biggs72} for an interesting counterpoint for the chromatic polynomials of graph ladders), which we can write in an even more compact form than eqs.~(\ref{expansionRelS0TnpourPetRho}) and (\ref{expansionRelS0SnpourPetRho}). If we can write --- assuming all the eigenvalues are distinct ---
\begin{equation}
{\rm Rel}_2^{(n)}(p,\rho) = \sum_{i \in \{0,+,-\}}^3 \alpha_i \, \lambda_i^n ,
\label{developpement general en lambda}
\end{equation}
the ordinary generating function ${\mathcal G}(x)$ defined by
\begin{equation}
{\mathcal G}(x) = \sum_{n=0}^{\infty} {\rm Rel}_2^{(n)}(p,\rho)  \, x^n
\label{developpement general de fonction generatrice}
\end{equation}
is easily calculated, since we have to sum geometric series :
\begin{equation}
{\mathcal G}(x) = \sum_{i \in \{0,+,-\}}^3 \frac{\alpha_i}{1 - \lambda_i \, x} = \frac{{\mathcal N}(x)}{{\mathcal D}(x)} .
\label{developpement general fonction generatrice}
\end{equation}
Of course, the summation can be performed too when $\alpha_i$ is a polynomial function of $n$. ${\mathcal G}(x)$ is a rational fraction of $x$ (see \cite{Stanley97} for a thorough discussion of the topics). Its denominator ${\mathcal D}(x) = \prod_i (1 - \lambda_i \, x)$ may be deduced very simply from the characteristic polynomial $P_{\rm carac}(x)$ of the unfactored transfer matrix by ${\mathcal D}(x) = (- x)^3 \, P_{\rm carac}(1/x)$. The determination of ${\mathcal G}(x)$, which encodes {\em all} the two-terminal reliabilities for the recursive family of graphs indexed by $n$, is actually quite simple: we only need to multiply the known leading terms of the expansion of ${\mathcal G}(x)$ by ${\mathcal D}(x)$ to see ${\mathcal N}(x)$ emerge.

For instance, in the case of ${\rm Rel}_2^{(S_0 \rightarrow T_n)}(p,\rho)$, we easily find ${\mathcal D}_T(x)$ from eq.~(\ref{Rel2S0Tnprhomatrice})
\begin{eqnarray}
{\mathcal D}_T(x) & = & 1 -  p \,\rho \, x \,\left( 2 + p \, \rho \, (1 - 2 \, p)  \right) \nonumber + p^2\,{\rho }^2\,x^2\,\left( 1 + p \, \rho \, (1 - 2 \, p) \, (2 - p \, \rho) \right) \nonumber \\[2mm]
& & - p^4\,{\rho }^4\,x^3\,\left( 1 - p\,\rho  \right) \,\left( 1 - 2\,p + p\,\rho  \right) ,
\label{denominateur pour S0 -> Tn avec p}
\end{eqnarray}
and deduce from eqs.~(\ref{RelS0T0pourPetRho})--(\ref{RelS0T2pourPetRho}) the simple result (mathematical softwares allowing symbolic calculus make such computations an easy task)
\begin{equation}
{\mathcal N}_T(x) = p \, \rho^2 \; \left( 1 - (1-p)\, p^2 \, \rho^2 \, x \right) .
\label{numerateur pour S0 -> Tn avec p}
\end{equation}

The preceding treatment works well too if all the $a_i$, $b_i$, and $c_i$ are replaced by $a$, $b$, and $c$, respectively, instead of by a unique value $p$. The final result reads, after simplification
\begin{eqnarray}
{\mathcal G}_T(a,b,c,\rho;x) & = & b \, \rho^2 \; \; \frac{{\mathcal N}_T(a,b,c,\rho;x)}{{\mathcal D}_T(a,b,c,\rho;x)} , \\[2mm]
{\mathcal N}_T(a,b,c,\rho;x) & = & 1 - a \, (1-b) \, c \, \rho^2 \, x , \\[2mm]
{\mathcal D}_T(a,b,c,\rho;x) & = & 1 - \big[a+c+a \, c \, \rho - 2 \, a \, b \, c \, \rho \big] \, \rho \, x \nonumber \\
& & + a \, c \, \big[1 + (1 - 2 \, b) \, (a+c) \, \rho - (1- a - c) \, b^2 \, \rho^2\big] \, \rho^2 \, x^2 \nonumber \\
& & - (1 - b \, \rho) \, (1 - 2 \, b + b \, \rho) \, a^2 \, c^2 \, \rho^4 \, x^3 .
\label{fonction generatrice pour S0 -> Tn avec a,b,c}
\end{eqnarray}
When $a$, $b$ and $c$ are not identical, no simple factorization of ${\mathcal D}_T$ occurs, so that we must use the roots of a polynomial of degree three in $x$, the literal expressions of which are not concise. However, the expression of the generating function makes it possible to calculate the two-terminal reliability by a mere partial fraction decomposition, which can be handled quite routinely.

In the case of ${\rm Rel}_2^{(S_0 \rightarrow S_n)}$, ${\mathcal G}_S(x) = {\mathcal N}_S(x)/{\mathcal D}_S(x)$, with ${\mathcal D}_S(x) = {\mathcal D}_T(x)$ and
\begin{eqnarray}
{\mathcal N}_S(x) & = & \rho - p \, \rho^2 \, \left(1 + p \, \rho \, (1 - 2 \, p) - p^2 \, \rho^2 \, (1-p) \right) \, x \nonumber \\
& & \hskip3mm + p^3 \, \rho^4 \, \left(1 - 2 \, p + p^2 \, \rho \, (2-\rho) \right) \, x^2 .
\label{numerateur pour S0 -> Sn avec p}
\end{eqnarray}
While this expression seems more complicated than eq.~(\ref{numerateur pour S0 -> Tn avec p}),  we can check that
\begin{equation}
{\mathcal G}_S(x)  - {\mathcal G}_T(x) = \frac{p \, \rho^2 \, (1-p \, \rho)^2 \, x}{1 - p \, \rho \, (1-p \, \rho) \, x} ,
\label{Difference GSn - GTn avec p}
\end{equation}
which could be directly deduced from eqs.~(\ref{expansionRelS0TnpourPetRho}) and (\ref{expansionRelS0SnpourPetRho}).

The generating function ${\mathcal G}_U$ for $S_0 \rightarrow U_n$ is given by
\begin{equation}
{\mathcal G}_U(x)  = p \, \rho^2 \, x \; \frac{(2-p) + p^2 \, \rho \, (1 - 2 \, \rho + p \, \rho) \, x }
{1 - p \, \rho \, \big(1 + 2 \, p \, (1-p) \, \rho\big) \, x + p^3 \, \rho^3 \, (1 - 2 \, p + p \, \rho) \, x^2} .
\label{GUn avec p}
\end{equation}
Here, the denominator is only a polynomial in $x$ of degree two, leading to the decomposition in eq.~(\ref{expansionRelS0UnpourPetRho}). We have included the case $n = 1$, for which the reliability is that of two links in parallel between $S_0$ and $U_1$, namely $\rho^2 \, (2 \, p - p^2)$; this has the advantage of simplifying the rational ${\mathcal G}_U(x)$, and making transparent the derivation of eq.~(\ref{expansionRelS0UnpourPetRho}).

\section{Zeros of the two-terminal reliability polynomials}
\label{Zeros}

One way to understand the structure of the different reliability polynomials is to study the locations of their zeros in the complex plane. Such a study has been fruitfully performed in the case of the chromatic polynomial \cite{Biggs72,Biggs01,Salas01}, most notably in the context of the four-color theorem. In the reliability context, some effort has been done to discover general properties for the all-terminal reliability ${\rm Rel}_A(p)$ \cite{Chari97,Colbourn93,Oxley02}, its main byproduct being the Brown-Colbourn conjecture \cite{BrownColbourn92}, according to which all the zeros are to be found in the region $|1-p| < 1$. Although valid for series-parallel graphs, this remarkable conjecture does not strictly hold in the general case (but not by far) \cite{Royle04}. As mentioned in the introduction, the all-reliability polynomial is linked to the Tutte polynomial, an invariant of the graph. It has also been studied extensively by Chang and Shrock for various recursive families of graphs \cite{Chang03}, who give the limiting curves where all zeros of the polynomials converge.

In this section, we want to locate the zeros of the two-terminal reliability polynomial, in order to see whether some insight may also be found in this case. Truly, this polynomial depends on the couple (source, terminal), but some structures could still be expected. We show indeed that the zeros tend to aggregate along sets of curves, which can substantially differ even for two seemingly similar ladder graphs.

\subsection{Calculation of the limiting curves: General results}

As $n$ grows, the number of zeros of the reliability polynomial in the complex plane increases. Because of the matrix transfer property, we have recursion relations between polynomials corresponding to successive values of $n$, and the solution looks like eq.~(\ref{FormeRelS0TnpourPetRho}). The general treatment of the problem has been done by Beraha, Kahane, and Weiss \cite{Beraha78}, but may be understood in the following, simplifying way: if the reliability polynomial has the form $\sum_i \alpha_i \, \lambda_i(p)^n$ (where $\lambda_i$ are the eigenvalues of the transfer matrix), then at large $n$, only the two eigenvalues of greater modules, say $\lambda_1$ and $\lambda_2$, will prevail, so that the reliability polynomial will vanish when $|\lambda_1(p)| = |\lambda_2(p)|$ (of course, it might be three or more eigenvalues of equal moduli; our simplification is sufficient here). This defines a set of curves in the complex plane, where all zeros should accumulate in the $n \rightarrow \infty$ limit. The interested reader should refer to the work of Salas and Sokal \cite{Salas01} for a very detailed discussion of the convergence to the limiting curves. This behavior would not be modified when one or more of the $\alpha_i$'s is a polynomial in $n$ \cite{Beraha78}.

In the following we limit ourselves to the perfect-nodes case ($\rho = 1$) and look for the limiting curves for the architectures considered in Section~\ref{Identical reliabilities}, because it is illustrative enough, and because all expressions are then simple and analytical. The trivial solution $p=0$ will not always be considered in the following.


\subsection{Calculation for the symmetrical ladder ($S_0 \rightarrow U_n$)}

Let us first consider the simplest case for ${\rm Rel}_2(S_0 \rightarrow U_n)$ given in eq.~(\ref{expansionRelS0UnpourPetRho}) because only two eigenvalues are present (see Fig.~\ref{Echelle Symetrique S0-Un}). We have to find $p$ such that $|x_+/x_-| = 1$, or
\begin{equation}
\left|\frac{x_+}{x_-}\right| = 1 = \left|\frac{1 + 2 \, p \, (1-p) + \sqrt{1+4 \, p^2 \, (1-p)^2}}{1 + 2 \, p \, (1-p) - \sqrt{1+4 \, p^2 \, (1-p)^2}}\right| .
\label{rapport lambdas Un}
\end{equation}
Introducing  $t = 2 \, p \, (1-p)$, we must find $t$ such that
\begin{equation}
\frac{1 + t + \sqrt{1+t^2}}{1 + t - \sqrt{1+t^2}} = e^{i \, \theta} \hskip3cm (- \pi \leq \theta \leq \pi) ,
\label{rapport lambdas Un avec t}
\end{equation}
which can be rewritten as
\begin{equation}
\sqrt{1+t^2} = i \, (1 + t) \, \tan \frac{\theta}{2} .
\label{rapport lambdas Un avec t 2}
\end{equation}
Squaring both sides of eq.~(\ref{rapport lambdas Un avec t 2}) implies
\begin{equation}
t^2 + 2 \, \sin^2 \frac{\theta}{2} + 1 = 0 ,
\label{rapport lambdas Un avec t 3}
\end{equation}
which yields two possible solutions, $t_{\pm}$, of modulus 1:
\begin{equation}
t_{\pm} = -\sin^2 \frac{\theta}{2} \pm \, i \, \sqrt{1 - \sin^4 \frac{\theta}{2}} .
\label{rapport lambdas Un avec t 4}
\end{equation}
Only the solutions that verify eq.~(\ref{rapport lambdas Un avec t 2}) must be kept ; they may be rewritten in condensed form
\begin{equation}
t = - e^{i \, \varphi} \hskip3cm \left(-\frac{\pi}{2} \leq \varphi \leq \frac{\pi}{2}\right) ,
\label{rapport lambdas Un avec t 5}
\end{equation}
which represents a half-circle (the real part of $t_{\pm}$ must be negative), so that the searched values of $p$ are finally
\begin{equation}
p = \frac{1 \pm \sqrt{1+2 \, e^{i \, \varphi}}}{2} \hskip3cm \left(-\frac{\pi}{2} \leq \varphi \leq \frac{\pi}{2}\right) ,
\label{p limites pour Un}
\end{equation}
It is then straightforward to give the coordinates of some particular points of the curves. On the real axis, the solutions for $\varphi = 0$ are $(1 \pm \sqrt{3})/2$, i.e., +1.366 and -0.366. The endpoints of the two curves are actually $\frac{1}{2} \, (1 \pm \sqrt{\frac{1 + \sqrt{5}}{2}}) \pm i \, \sqrt{\frac{\sqrt{5} - 1}{8}}$, or $1.136 \pm 0.393 \, i$ and $-0.136 \pm 0.393 \, i$.

For $n = 10$, we have
\begin{eqnarray}
{\rm Rel}_2(S_0 \rightarrow U_{10}) & = & p^{10}\,\big( 2 + 18\,p + 68\,p^2 + 100\,p^3 - 134\,p^4 - 746\,p^5 - 648\,p^6 \nonumber \\
& &  + 1824\,p^7 + 3818\,p^8 - 2354\,p^9 - 10861\,p^{10} + 2586\,p^{11} \nonumber \\
& &  + 23080\,p^{12} - 7904\,p^{13} - 48624\,p^{14} + 79008\,p^{15} \nonumber \\
& &  - 58432\,p^{16} + 24064\,p^{17} - 5376\,p^{18} + 512\,p^{19} \big) .
\label{S0 -> U10}
\end{eqnarray}

\vskip0.6cm
\begin{figure}[thb]
\hskip2cm
\includegraphics[width=0.75\linewidth]{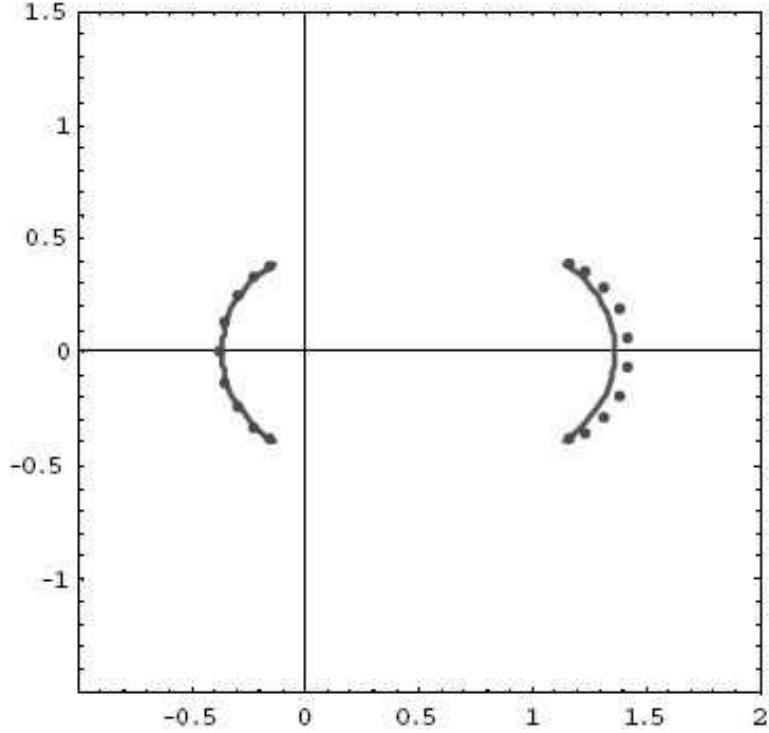}
\vskip0.3cm
\caption{Location of the zeros of ${\rm Rel}_2^{(S_0 \rightarrow U_n)}(p)$: the dots
corresponds to $n = 10$ and eq.~(\ref{S0 -> U10}), the continuous curves to the limit $n
\rightarrow \infty$ given by eq.~(\ref{p limites pour Un}). The trivial solution $p=0$ is not
displayed.} \label{Zeros pour Un}
\end{figure}
\vskip0.6cm

We see on Fig.~\ref{Zeros pour Un} that for $n=10$ already, the zeros are located very close to the
two well-separated arcs of quasi-circular shape defined by eq.~(\ref{p limites pour Un}). Note that
with respect to the diagrams by Chang and Shrock for the all-terminal reliability polynomials
calculated for several recursive families of graphs \cite{Chang03}, we now have accumulation points
in the region ${\rm Re}(p) < 0$. When $\rho \neq 1$, the global appearance is unchanged, even if
the zeros move away from the origin. When $\rho$ approaches zero, the structure distorts itself to
look like a circle of radius of the order of $(2 \, \rho)^{-1/2}$ centered at $(\frac{1}{2},0)$.
The convergence of the roots to the limiting curve follows the general expression of Salas and
Sokal \cite{Salas01}.

\subsection{Calculation for ${\rm Rel}_2(S_0 \rightarrow S_n)$ and ${\rm Rel}_2(S_0 \rightarrow T_n)$}

In these configurations, we now have three eigenvalues. We must therefore consider $|x_{\pm}/x_0|$ in addition to $|x_+/x_-|$, calculated in the preceding section. We must find all the $p$'s such that
\begin{equation}
\frac{x_{\pm}}{x_0} = \frac{1 + 2 \, p \, (1-p) \pm \sqrt{1+4 \, p^2 \, (1-p)^2}}{2 \, (1-p)} = e^{i \, \theta} \hskip1cm (- \pi \leq \theta \leq \pi) ,
\label{rapport lambdas Tn}
\end{equation}
so that
\begin{equation}
Z = \pm \sqrt{1+4 \, p^2 \, (1-p)^2} = 2 \, (1-p) \, e^{i \, \theta}  - \left( 1 + 2 \, p \, (1-p) \right) .
\label{rapport lambdas Tn 2}
\end{equation}
Here again, we square both sides of eq.~(\ref{rapport lambdas Tn 2}), keeping in mind we will have to check that the real part of the right-hand-side of eq.~(\ref{rapport lambdas Tn 2}) must comply with our determination of the square root (our angle argument being taken between $- \pi$ and $+ \pi$, we choose ${\rm Re} \sqrt{\cdots} \geq 0$). After simplification, we get
\begin{equation}
1 + 2 \, p \, (1-p) = (1 - p) \, e^{i \, \theta} + p \, e^{- i \, \theta} .
\label{rapport lambdas Tn 3}
\end{equation}
This equation has two possible solutions, namely
\begin{equation}
p_{\pm} = \frac{1}{2} \left( 1 + i \, \sin \theta \pm \sqrt{1 + (1 - \cos \theta)^2}\right) .
\label{rapport lambdas Tn 4}
\end{equation}
Which values of $\theta$ satisfy with eq.~(\ref{rapport lambdas Tn 2}) ? Using eqs.~(\ref{rapport lambdas Tn 2}) and (\ref{rapport lambdas Tn 3}), we find
\begin{equation}
Z = (1 - p_{\pm}) \, e^{i \, \theta} - p_{\pm} \, e^{- i \, \theta} = \mp \cos \theta \, \sqrt{1 + (1 - \cos \theta)^2} + i \, \sin \theta \, (1 - \cos \theta) ,
\label{rapport lambdas Tn 4}
\end{equation}
so that ${\rm Re}(Z)$ and ${\rm Re}(\mp \cos \theta)$ have the same sign. Since for $x_+/x_0$, we must have ${\rm Re}(Z) \geq 0$ and thus ${\rm Re}(\mp \cos \theta) \geq 0$, we find
\begin{eqnarray}
p[|x_+/x_0|=1] & = & \left\{
\begin{array}{l}
p_+  \hskip1cm ( \frac{\pi}{2} \leq |\theta| \leq \pi) \\
p_-  \hskip1cm ( 0 \leq |\theta| \leq \frac{\pi}{2})
\end{array}
\right. ,
\\
p[|x_-/x_0|=1] & = & \left\{
\begin{array}{l}
p_-  \hskip1cm ( \frac{\pi}{2} \leq |\theta| \leq \pi) \\
p_+  \hskip1cm ( 0 \leq |\theta| \leq \frac{\pi}{2})
\end{array}
\right. .
\label{rapport lambdas Tn 5}
\end{eqnarray}
We also have to check, when considering $p[|x_+/x_0|=1]$ for instance, that the third eigenvalue $x_-$ has a smaller modulus. Consequently, using eqs.~(\ref{rapport lambdas Tn 3}) and (\ref{rapport lambdas Tn 4}),
\begin{equation}
\left|\frac{x_+}{x_-}\right| = \left|\frac{(1 - p) \, e^{i \, \theta} + p \, e^{- i \, \theta} + ((1 - p) \, e^{i \, \theta} - p \, e^{- i \, \theta})}{(1 - p) \, e^{i \, \theta} + p \, e^{- i \, \theta} - ((1 - p) \, e^{i \, \theta} - p \, e^{- i \, \theta})} \right| = \left|\frac{1-p}{p}\right| \geq 1 ,
\label{rapport lambdas Tn 6}
\end{equation}
and therefore that ${\rm Re}(p) \leq \frac{1}{2}$, where $p$ stands for $p_{\pm}$. We deduce from eq.~(\ref{rapport lambdas Tn 4}) that the $p_+$ solution is to be discarded. The same procedure for $p[|x_-/x_0|=1]$ shows that only $p_-$ should be considered.

\begin{figure}[thb]
\hskip2cm
\includegraphics[width=0.75\linewidth]{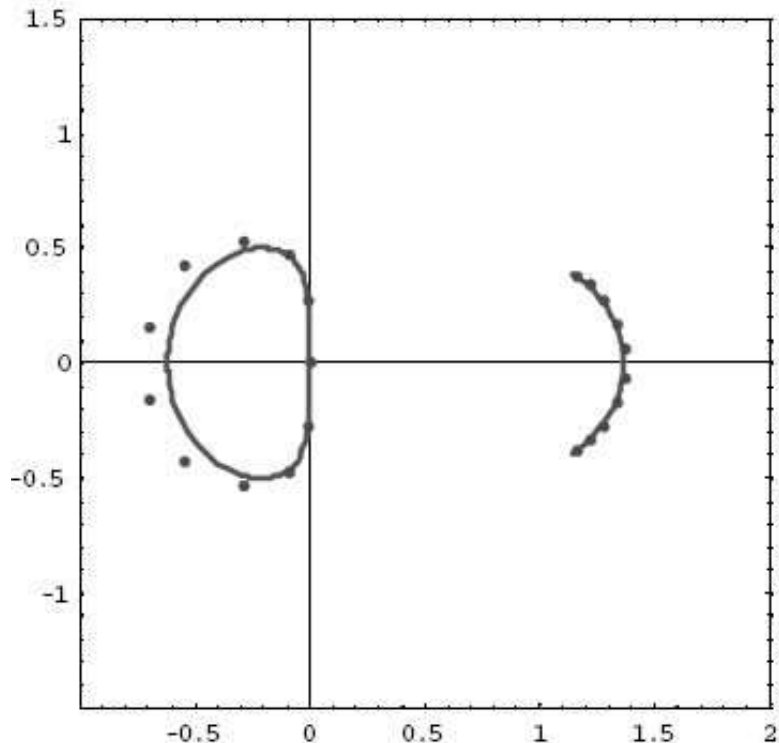}
\vskip0.3cm
\caption{Location of the zeros of ${\rm Rel}_2^{(S_0 \rightarrow T_n)}(p)$: the dots corresponds to $n = 10$ (eq.~(\ref{S0 -> T10})), the continuous curves to the limit $n \rightarrow \infty$ given in eq.~(\ref{courbes limites zeros pour Tn}).}
\label{Zeros pour Tn}
\end{figure}

Finally, we must assess the modulus of $x_0$ along the curves given by eq.~(\ref{p limites pour Un}). Here, the situation is slightly more complicated, and we checked numerically that the $p_-$ solution of eq.~(\ref{p limites pour Un}) should be discarded.
The final result
\begin{equation}
p  =
\left\{
\begin{array}{lll}
\displaystyle \frac{1 + \sqrt{1+2 \, e^{i \, \varphi}}}{2} & \hskip1cm & (-\frac{\pi}{2} \leq \varphi \leq \frac{\pi}{2}) \\
\frac{1}{2} \left( 1 - \sqrt{1 + (1 - \cos \theta)^2} + i \, \sin \theta \right) & \hskip1cm & (-\pi \leq \theta \leq \pi)
\end{array}
\right.
\label{courbes limites zeros pour Tn}
\end{equation}
is displayed in Fig.~\ref{Zeros pour Tn}.

\vskip0.6cm
\vskip0.6cm

When $n=10$, we have
\begin{eqnarray}
{\rm Rel}_2(S_0 \rightarrow T_{10}) & = & p^{11} \, \big( 11 + 155\,p^2 - 99\,p^3 + 40\,p^4 - 907\,p^5 - 296\,p^6 \nonumber \\
& &  + 1448\,p^7 + 3121\,p^8 - 1102\,p^9 -  7989\,p^{10} - 1747\,p^{11} \nonumber \\
& &  + 14806\,p^{12} + 4776\,p^{13} - 24168\,p^{14} + 176\,p^{15} + 35072\,p^{16} \nonumber \\
& &  - 38016\,p^{17} + 19072\,p^{18} - 4864\,p^{19} + 512\,p^{20} \big) .
\label{S0 -> T10}
\end{eqnarray}
We have plotted in Fig.~\ref{Zeros pour Tn} the zeros for $n=10$ and the $n \rightarrow \infty$ limit. Note that the asymptotic limit seems to be reached faster in the ${\rm Re}(p) > 0$ half-plane. The new curve endpoints of Fig.~\ref{Zeros pour Tn} are $\frac{1}{2} \, (1 - \sqrt{2} \pm i) \approx -0.207 \pm 0.5 \, i$, and  $\frac{1}{2} \, (1 - \sqrt{5}) \approx -0.618$. For $\rho \neq 1$, the structure of the zeros spreads out from the origin, as in the preceding subsection.

We want to emphasize that while the structure of the zeros in Figs.~\ref{Zeros pour Un} and \ref{Zeros pour Tn} is similar in the ${\rm Re}(p) > 0$ half-plane --- the $n \rightarrow \infty$ limit is identical --- this is certainly not the case in the ${\rm Re}(p) < 0$ half-plane, even though the two graphs differ by two links only. Inferences from these structures of zeros should therefore be considered cautiously.

\section{Sensitivity}
\label{Sensibilite}

An important parameter appearing in reliability studies is the sensitivity $s_i$ \cite{Rubino91}, which measures the influence of a particular equipment $i$ (node or edge) on the global reliability; the sensitivity analysis is akin to the various importance criteria compiled by Henley et Kumamoto \cite{Henley91}. It is usually defined as the derivative of the total --- be it two- or all-terminal --- reliability with the individual reliability $p_i$
\begin{equation}
s_i = \frac{\partial {\rm Rel}_{2,A}}{\partial p_i} = {\rm Rel}_{2,A}(p_i = 1) - {\rm Rel}_{2,A}(p_i = 0) .
\label{U 4 pannes pour Sn et Tn}
\end{equation}
The two definitions are equivalent, because ${\rm Rel}_{2,A}$ is an affine function of all the $p_i$'s. The calculation can be performed quite easily, since the desired $p_i$ appears in {\em only one} of the transfer matrices of Section \ref{Main calculation}. For instance, let us consider the sensitivity of the successive ``rungs'' of the ladder --- a similar work could be done for the nodes, without difficulty. In the following, we restrict ourselves to the case ${\rm Rel}_2(S_0 \rightarrow T_n)$, with perfect nodes and identical edge reliabilities $p$, and want to assess the influence of the edge location to the overall performance of the network.

From eq.~(\ref{transfermatrixS0Tn}), taking the derivative of a matrix with respect to a given $b_{n_0}$ is not difficult, since each of its elements is an affine function of $b_{n_0}$. A new matrix appears in the problem, and when all edge reliabilities are set to $p$ (we take $\rho = 1$ for the sake of simplicity), we have
\begin{eqnarray}
s_{b_{n_0}} & = &
(0,1,0) \, \cdot \, \left(
\begin{array}{c|c|c}
p & p^3 & p^3 \\ \hline
p & p & p^2 \\ \hline
- p^2 & - p^3 & p^2 \, (1 - 2 \, p)
\end{array}
\right)^{n-n_0} \, \cdot \, \left(
\begin{array}{c|c|c}
0 & p^2 & p^2 \\ \hline
1 & 0 & p \\ \hline
- p & - p^2 & - 2 \, p^2
\end{array}
\right) \, \nonumber \\
& & \cdot \,
 \left(
\begin{array}{c|c|c}
p & p^3 & p^3 \\ \hline
p & p & p^2 \\ \hline
- p^2 & - p^3 & p^2 \, (1 - 2 \, p)
\end{array}
\right)^{n_0} \, \cdot \,
 \left(
\begin{array}{c} 1 \\ 0 \\ 0 \end{array}
\right) .
\label{sensitivity bn0}
\end{eqnarray}

For symmetry reasons, the contributions of $b_i$ and $b_{6-i}$ are identical. For $n=6$,
\begin{eqnarray}
s_{b_0} & = & (1 - p) \, p^6\,( 1 + p + 15\,p^2 + 4\,p^3 - 18\,p^4 - 55\,p^5 + p^6 + 116\,p^7 \nonumber \\
& &  + 24\,p^8 - 200\,p^9 + 144\,p^{10} - 32\,p^{11} ) , \\
s_{b_1} & = & (1 - p)^2\,p^6\,( 1 + 2\,p + 16\,p^2 + 15\,p^3 - 11\,p^4 - 60\,p^5 - 28\,p^6 \nonumber \\
& &  + 92\,p^7 + 40\,p^8 - 96\,p^9 + 32\,p^{10} ) , \\
s_{b_2} & = & \nonumber (1 - p)^2\,p^6\,( 1 + 2\,p + 16\,p^2 + 14\,p^3 - 14\,p^4 - 61\,p^5 - 20\,p^6 \nonumber \\
& &  + 88\,p^7 + 40\,p^8 - 96\,p^9 + 32\,p^{10} ) , \\
s_{b_3} & = & (1 - p)^2\,p^6\,( 1 + 2\,p + 16\,p^2 + 14\,p^3 - 15\,p^4 - 60\,p^5 - 20\,p^6 \nonumber \\
& &  + 88\,p^7 + 40\,p^8 - 96\,p^9 + 32\,p^{10} ) .
\label{sensibilite S0->T6}
\end{eqnarray}

\vskip0.6cm
\begin{figure}[thb]
\hskip2cm
\includegraphics[width=0.75\linewidth]{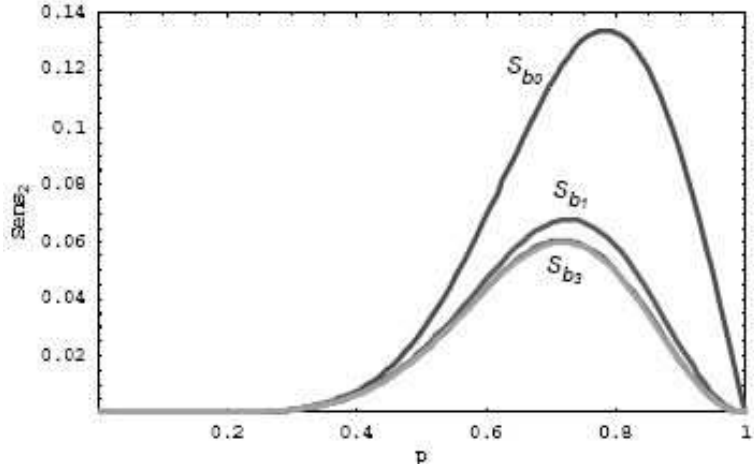}
\vskip0.3cm
\caption{Variation of the sensitivity ${\rm Sens}_2$ of the ladder's rungs, depending on their location, in the ${\rm Rel}_2(S_0 \rightarrow T_6)$ case.}
\label{sensibilites pour S0->T6}
\end{figure}
\vskip0.6cm

Figure \ref{sensibilites pour S0->T6} shows that these sensitivities vary with $p$ and that they uniformly decrease from the ladder's ends to its center. Note that only a few coefficients of the polynomials differ. If the length of the ladder increases, the maximum sensitivity will actually decrease in magnitude, and be located at higher values of $p$. Besides, if $n$ is odd, then the minima are shared by the two central rungs, i.e., $n_0 = (n \pm 1)/2$.

\begin{figure}[thb]
\hskip1.5cm
\includegraphics[width=0.75\linewidth]{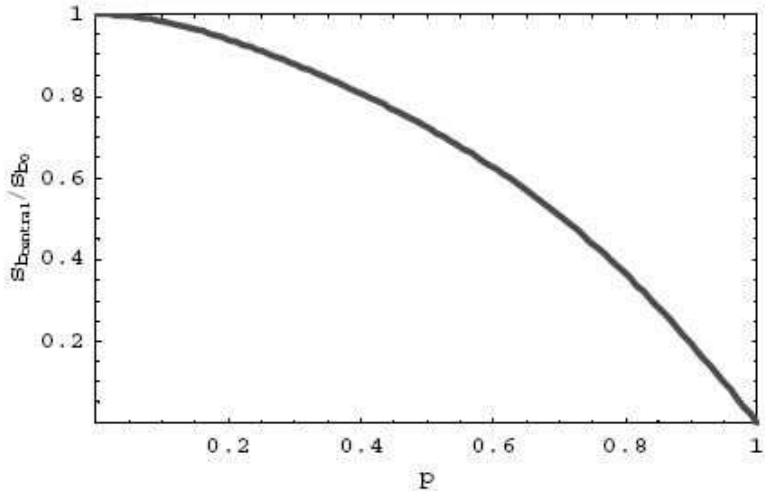}
\vskip0.3cm
\caption{Ratio $s_{b_{\rm central}}/s_{b_0}$ for ${\rm Rel}_2(S_0 \rightarrow T_{2 n})$ in the large $n$ limit.}
\label{RapportSensibilites}
\end{figure}

Because of the form of eq.~(\ref{sensitivity bn0}), the final result is expected to be a sum of different power-law contributions, with products of two of the different eigenvalues already observed for the reliability. The easier way is then to calculate the generating functions in a $S_0 \rightarrow T_{2 \, n}$ ladder, for both $b_0$ and $b_n \equiv b_{\rm central}$. We finally get ${\mathcal G}_{\rm sens}(b_{i}) = \sum_{n=0} \, s_{b_i}(S_0 \rightarrow T_{2 \, n}) \, x^n$ with ${\mathcal G}_{\rm sens} = {\mathcal N}_{\rm sens}/{\mathcal D}_{\rm sens}$:

\begin{eqnarray}
{\mathcal N}_{\rm sens}(b_{\rm central}) & = & 1 - p^2\,(1 + p + 3\,p^2 - 6\,p^3 + 2\,p^4)\, x \nonumber \\
& & + \left(1 - p\right) \,p^5\,( 1 + 3\,p - 3\,p^2 - 2\,p^3 + 2\,p^4) \, x^2 \nonumber \\
& & - p^9 \, (1-p)^3 \, x^3 , \label{N pour bcentral} \\
{\mathcal D}_{\rm sens}(b_{\rm central}) & = & \Big( 1 - p^2
\, (1-p)^2 \, x\Big) \, \Big( 1 - p^3 \, (1-p) \, x\Big) \nonumber \\
& & \hskip-1cm \times \Big( 1 - p^2\,( 1 + 2\,p + 2\,p^2 -
8\,p^3 + 4\,p^4 ) \, x + p^6 \, (1-p)^2 \, x^2\Big) .
\label{D pour bcentral}
\end{eqnarray}
whereas
\begin{equation}
{\mathcal G}_{\rm sens}(b_{0}) = \frac{1 - p^2\,(1 + 3\,p^2 - 5\,p^3 + 2\,p^4 ) \,x + (2-p) \, p^6 \, (1-p)^2 \,
x^2}{\big(1 - p^2 \, (1-p)^2 \, x\big) \, \Big(1 - p^2\,( 1 + 2\,p + 2\,p^2 - 8\,p^3 + 4\,p^4 ) \, x + p^6 \, (1-p)^2 \, x^2\Big)} .
\label{G pour b0}
\end{equation}
The partial fraction decompositions of ${\mathcal G}_{\rm sens}(b_{0})$ and ${\mathcal G}_{\rm sens}(b_{\rm central})$ lead to
\begin{eqnarray}
s_{b_0} & = & \frac{1}{2} \, p^{2 \, n} \, \left\{ (1-p)^{2 \, n} \, + \frac{x_+^{2 \, n}}{2} \, \left(1 + \frac{1-2 \, p^2}{\sqrt{1 + 4 \, p^2 \, (1-p)^2}}  \right) \right. \nonumber \\
& & \hskip2cm \left. + \frac{x_-^{2 \, n}}{2} \, \left(1 - \frac{1-2 \, p^2}{\sqrt{1 + 4 \, p^2 \, (1-p)^2}}  \right) \right\}
\label{sensibilite b0}
\end{eqnarray}
and
\begin{eqnarray}
s_{b_{\rm central}} & = & \frac{1}{2} \, p^{2 \, n} \, \left\{ (1-p)^{2 \, n} \, + \frac{4 \, p^2 \, (1-p)}{1 + 4 \, p^2 \, (1-p)^2} \, p^n \, (1-p)^n \right. \nonumber\\
& & \hskip1.2cm \left. + \frac{x_+^{2 \, n}}{4} \, \left(1 + \frac{1-2 \, p^2}{\sqrt{1 + 4 \, p^2 \, (1-p)^2}}  \right)^2 \right. \nonumber\\
& & \hskip1.2cm \left. + \frac{x_-^{2 \, n}}{4} \, \left(1 - \frac{1-2 \, p^2}{\sqrt{1 + 4 \, p^2 \, (1-p)^2}}  \right)^2 \right\} ,
\label{sensibilite bcentral}
\end{eqnarray}
so that the ratio $s_{b_{\rm central}}/s_{b_0}$, in the large $n$ limit, is given by
\begin{equation}
\frac{s_{b_{\rm central}}}{s_{b_0}} \longrightarrow  \frac{1}{2} \, \left(1 + \frac{1-2 \, p^2}{\sqrt{1 + 4 \, p^2 \, (1-p)^2}}  \right) ,
\label{rapport sensitivites}
\end{equation}
as displayed in Fig.~\ref{RapportSensibilites} (for $p = 1-q$ close to 1, this ratio expands as $2 \, q - 6 \, q^3 + 8 \, q^4 + \cdots$).

\vskip0.6cm
\vskip0.6cm
Actually, this limit is reached very quickly, since for the $S_0 \rightarrow T_6$ ladder, the ratio $s_{b_3}/s_{b_0}$ is barely distinguishable from the asymptotic limit in the entire range $0 \leq p \leq 1$, which is displayed in Fig.~\ref{RapportSensibilites}. It also shows that the overall reliability is more sensitive to ``access'' rungs located at the edges of the ladder network than to those in the middle, even if all edge reliabilities are identical.

\section{All-terminal reliability}
\label{All-terminal reliability}

As mentioned in the introduction, the all-terminal reliability is another useful measure of the network availability, namely the probability that all nodes are connected. We refer the reader to the work of Chang and Shrock \cite{Chang03} for the explicit expressions of ${\rm Rel}_A$ for various recursive families of graphs, among which our simple ladder graph, when all edges have the same reliability $p$. For the sake of completeness, we show that their result may be slightly generalized for edges with distinct reliabilities. Here again, the final, analytical expression can be written in a concise form using transfer matrices. In the context of graph theory, this can be viewed as the factorization of a particular value of the multi-variate Tutte polynomial considered by Wu \cite{Wu78} and Sokal \cite{Sokal05}.

\vskip0.6cm
\begin{figure}[thb]
\hskip0cm
\includegraphics[width=0.75\linewidth]{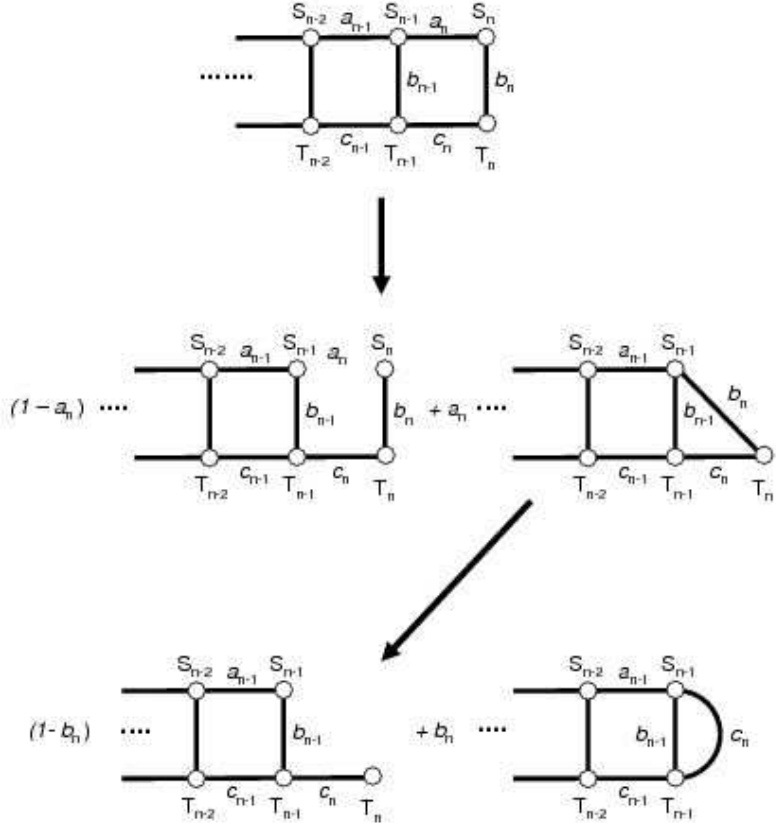}
\vskip0.3cm
\caption{Successive deleting/contracting operations lead to a recursion relation between ${\mathcal R}_n$ and ${\mathcal R}_{n-1}$ (see text).}
\label{Decomposition pour all-terminal}
\end{figure}
\vskip0.6cm

We expect the all-terminal reliability ${\mathcal R}_n$ for the simple ladder diagram of Fig.~\ref{DeuxEchelles} should exhibit the same behavior, with a transfer matrix depending on $a_n$, $b_n$, and $c_n$, and a generating function that is a rational fraction of $p$ (the common reliability of links) and $x$. Actually, the calculation of the all-terminal reliability may be somewhat easier because all nodes can be considered perfect without loss of generality \cite{Colbourn87}. We can then use the usual factoring decomposition to establish a relationship between ${\mathcal R}_n$ and ${\mathcal R}_{n-1}$. From Fig.~\ref{Decomposition pour all-terminal}, we easily find that

\begin{equation}
{\mathcal R}_n = \left[ (1 - a_n) \, b_n + a_n \, (1 - b_n) \right] \, c_n \, {\mathcal R}_{n-1} + a_n \, b_n \,
{\mathcal R}_{n-1} (b_{n-1} \rightarrow b_{n-1} \, // \, c_n) .
\label{recurrence pour Rn}
\end{equation}

Using the ansatz ${\mathcal R}_n = \alpha_n \, (a_n + b_n) + \beta_n \, a_n \, b_n$, we get
\begin{equation}
\left(
\begin{array}{l}
\alpha_n \\
\beta_n
\end{array}
\right) = \left(
\begin{array}{l|l}
(a_{n-1} + b_{n-1}) \, c_n & a_{n-1} \, b_{n-1} \, c_n \\ \hline
(a_{n-1} + b_{n-1}) \, (1 - 2 \, c_n) + c_n \, (1 - 2 \, b_{n-1}) & a_{n-1} \, \big( c_n + b_{n-1} \, (1
- 3 \, c_n) \big)
\end{array}
\right) \; \left(
\begin{array}{l}
\alpha_{n-1} \\
\beta_{n-1}
\end{array}
\right) ,
\label{matrice de transfert pour Rn}
\end{equation}
where actually $\alpha_0 = 1$ and $\beta_0 = 0$. Calling $\widehat{M}_n$ the matrix appearing in eq.~(\ref{matrice de transfert pour Rn}), we obtain (with the condition $a_0 = 0$ in $\widehat{M}_1$)
\begin{equation}
{\mathcal R}_n = (a_n + b_n, a_n \, b_n) \; \cdot \; \widehat{M}_n \; \cdot \; \widehat{M}_{n-1} \; \cdots  \; \widehat{M}_1 \;  \cdot \; \left(
\begin{array}{l}
1 \\
0
\end{array}
\right) .
\label{expression de Rn}
\end{equation}
When all the reliabilities are equal to $p$, the transfer matrix is simply
\begin{equation}
M_R = \left(
\begin{array}{l|l}
2 \, p^2 &  p^3 \\ \hline
p \, (3 - 5 \, p) & p^2 \, (2 - 3 \, p)
\end{array}
\right) .
\label{matrice transfert pour Rn avec p}
\end{equation}

The eigenvalues of $M_R$, $\displaystyle \zeta_{\pm} = p^2 \; \frac{4 - 3 \, p
\pm \sqrt{12 - 20 \, p + 9 \, p^2}}{2}$, are easily obtained, and finally
\begin{equation}
{\mathcal R}_n = \frac{1}{p \, \sqrt{12 - 20 \, p + 9 \, p^2}} \; \left(\zeta_{+}^{n+1} - \zeta_{-}^{n+1} \right)
\label{Rn final}
\end{equation}
and
\begin{equation}
G_R(x) = \frac{p}{1 - p^2 \, (4 - 3 \, p) \, x + p^4 \, (1-p) \, x^2} .
\label{GRn final}
\end{equation}
The last results were obtained in \cite{Chang03}: the polynomials aggregate at the curve defined by $p = 1+ \frac{1}{3} \, e^{i \, \varphi}$, with $\cos \varphi \geq \frac{1}{3}$. It is worth stressing that the new scale $\zeta_+$ for the all-terminal reliability is different from the eigenvalues found in the two-terminal study, but that a power-law behavior is still the rule.

\section{Conclusion and perspectives}
\label{Perspectives}

We have found the solution of the two- and all-terminal reliabilities for a simple ladder graph, which corresponds nonetheless to realistic network architectures, especially in telecommunication networks for IP transport, but not only. Node and edge failures are put on an equal footing, and the simple formulae relying on transfer matrices may be directly implemented, even in a worksheet application. When identical reliabilities $p$ and $\rho$ are assumed for edges and nodes, respectively, we have given the analytical solution of the two-terminal reliability, for which only numerical determinations were previously available. We have shown that, while it may be useful to study the location of the zeros of the two-terminal reliability polynomial, it may strongly depend on the existence of a limited number of particular edges, especially in the ${\rm Re}(p) < 0$ region. We have also given the generating functions of the reliability polynomials, and provided a glimpse of sensitivity studies, which could be performed straightforwardly.

Although we used a delta-star transformation to solve a particular family of graphs, the present work may clearly be extended in several directions, which we outline in the following, and which we shall develop elsewhere.

\subsection{Brecht-Colbourn ladder}

Another simple ladder graph is the Brecht-Colbourn ladder \cite{BrechtColbourn86,Prekopa91}, which has been considered as a case study, both for the two- and all-terminal reliability, in order to evaluate the quality of bounds for the reliability polynomial in the case of perfect nodes. It has been also brought forward in a special case of broadcasting network \cite{Graver05}. Using the same delta-star transformation, the application of which is slightly more tricky, we have been able to solve this problem exactly, where the intermediate nodes have a connectivity degree of 4 instead of 3. The complete results --- the transfer matrix is then of dimension four for imperfect nodes, but still of dimension three for perfect nodes; a few eigenvalues may be complex --- will be given elsewhere \cite{Tanguy06b}, along with results for a generalized fan \cite{Aggarwal75,Neufeld85}.

\subsection{More general recursive families of graphs}

Even though the delta-star transformation may not apply successfully to all networks, it seems quite clear that a similar decomposition through transfer matrices of two- and all-terminal reliabilities should exist too for a ladder of $K_4$ graphs, or ladders of greater width. In order to make such calculations useful for applications, imperfect nodes as well as imperfect edges should be considered. All is needed is a recursion relation between successive graphs, when one ``elementary cell'' is added. This implies a new expression for the deletion-contraction theorem, in which the --- most useful --- linearity with respect to all individual edge or node reliabilities must be preserved, whereas to our knowledge the edge reliabilities have been renormalized to account for the unreliability of the nodes they connect \cite{Theologou91,Torrieri94}. This new expression will be given elsewhere \cite{Tanguy06c}, along with an application to other recursive families of graphs ($K_4$ ladders and $K_3$ cylinders).



What should we expect ? Basically the same kind of behavior as detailed in the present work, with a factorization of the reliability in terms of transfer matrices, the dimension of which may substantially increase to reflect the interplay of different edges/nodes in the overall reliability, and the underlying algebraic structure of the graph. It should be clear, however, that all graphs might not behave that way \cite{Biggs72}. For instance, the generating function of the two-terminal reliability for the complete graphs $K_n$, for which we give a recursion relation in the case of imperfect nodes and edges in the appendix, might not be a rational fraction.

\subsection{Failure frequency of systems}

Another performance index for networks, mostly investigated in the context of power distribution
systems \cite{Hayashi91,Shi81,SinghBillinton74}, is the failure frequency of a connection.
Calculations can readily be performed when the assumption is made that all equipments fail at a
constant failure rate $\lambda_e$ and brought back to operation via a constant repair rate $\mu_e$ (and $\displaystyle p_e = \frac{\mu_e}{\mu_e + \lambda_e}$). The failure frequency is then the sum over all equipments of the products of the individual failure rate, reliability and sensitivity (as defined in Section \ref{Sensibilite}). The present results
clearly indicate that such calculations are made very simple by the transfer matrix formulation,
even for extended networks. We shall present various examples in a forthcoming paper \cite{Tanguy06d}.

\subsection{Reliability incertitude}

In recent years, Coit and collaborators \cite{Coit97,Coit04,Jin01} have considered the possible influence of the reliability uncertainty to the determination of an ``optimal network architecture'', and have mostly dealt with series-parallel architectures.  In the context of meshed networks, and recalling our assumption of statistically independent failures for the network constituents, we could safely replace in the final expression of the reliability each $p_e$ or $p_n$ by the random variable $P_e$ or $P_n$ of which it is the mean value. We can thus straightforwardly assess the influence of the full statistical properties of $P_e$ or $P_n$ (variance, skewness, kurtosis, etc.). This issue will be reported elsewhere. We can also replace, for non-reparable systems, each $p_e$ by the often used $\exp (- \lambda_e \, t)$.

\subsection{Bounds, combinatorics, etc.}

There are obviously many directions in which this work may be useful in the determination of bounds. The first one is to use our present results on simple ladder graphs, or their future extensions \cite{Tanguy06b,Tanguy06c}, as possible upper or lower bounds to more complex graphs. If a particular graph looks like a special instance of a recursive family of graphs, we expect the generating function to be a rational fraction again. The dimension of the corresponding transfer matrix may be probed by trying to find recursion relations between successive reliability polynomials. Of course, if the dimension of the corresponding transfer matrix is large, the degree of the numerator and denominator of the fraction may be too large for a complete solution to be obtained easily. Even so, the knowledge that the generating function is rational may be an indication that an {\em approximate} generating function might be quite useful, since Pad\'{e} approximants \cite{Baker96} are known for their devilish knack of getting very close to the exact function \cite{NumericalRecipes}. This suggests that we may profitably consider rational fractions --- deduced from approximate generating functions --- for bounds, instead of exclusively relying on polynomials. This will be demonstrated in detail elsewhere \cite{Tanguy06e}.

Our calculations may also provide some information on some combinatorial issues, such as the enumeration of self-avoiding walks on lattices of restricted width. Finally, the exact results found for classes of arbitrarily large networks may prove useful for testing different algorithms (Monte Carlo, genetic, OBDD, etc.) in numerically exacting configurations, where edge and node unreliabilities have to be taken into account.







\section*{Acknowledgments}
\label{Acknowledgments}

I am indebted to \'{E}lisabeth Didelet for giving me the opportunity to study reliability issues in
depth, and to Jo\"{e}l Le Meur, Fran\c{c}ois Gallant, Veluppillai Chandrakumar, Adam Ouorou, and
\'{E}ric Gourdin for encouragement and support. G\'{e}rard Cohen provided important information. I
would also like to thank Christine Leroy, Wojtek Bigos, Annalisa Morea, Antoine Pardigon, and Annie Druault-Vicard for
useful discussions.

\vfill
\eject

\appendix

\section{Two-terminal reliability for a complete graph with imperfect nodes}
\label{Complete graph}

The exact all-terminal reliability polynomial $A_n = {\rm Rel}_A(K_n)$ for the
complete graph $K_n$ with perfect nodes and identical edges has been known for
decades, as well as the corresponding two-terminal reliability polynomial $T_n
= {\rm Rel}_2(K_n)$ (\cite{Colbourn87}, pp. 33--34). They may be found by
recurrence, starting with $A_1 = 1$:
\begin{equation}
A_n = 1 - \sum_{j=1}^{n-1} \, \left(
\begin{array}{c}
n-1\\
j-1
\end{array} \right)
\, A_j \, (1-p)^{j \, (n-j)} \label{An}
\end{equation}
and
\begin{equation}
T_n = \sum_{j=2}^{n} \, \left(
\begin{array}{c}
n-2\\
j-2
\end{array} \right)
\, A_j \, (1-p)^{j \, (n-j)} .
\end{equation}
Taking imperfect nodes with identical reliability $\rho$ into account does not
substantially change the all-terminal reliability, since one merely has to
multiply the result of eq.~(\ref{An}) by $\rho$ to the power of the number of
operating nodes \cite{Colbourn87}. The situation is quite different for $K$-terminal
reliability, and in particular for two-terminal reliability, since all the
possible paths between the source and the destination may visit different
numbers of nodes. This can be very useful since the complete graph
configuration represents the highest possible upper bound, because all its
nodes are connected.

Defining $T_n(p,\rho)$ the two-terminal reliability polynomial with imperfect
edges and nodes --- so as to keep the notation of Colbourn \cite{Colbourn87}, we have
\begin{eqnarray}
T_2(p,\rho) & = & p \, \rho^2 , \nonumber \\
T_3(p,\rho) & = & p\,{\rho }^2 + p^2\,{\rho }^3 - p^3\,{\rho }^3 , \\
T_4(p,\rho) & = & p\,{\rho }^2 + 2\,p^2\,{\rho }^3 - 7\,p^4\,{\rho }^4 + 7\,p^5\,{\rho }^4 - 2\,p^6\,{\rho }^4 + p^3\,\left( -2\,{\rho }^3 + 2\,{\rho }^4 \right) ,
\nonumber
\end{eqnarray}
and more generally
\begin{equation}
\frac{T_n(p,\rho)}{\rho^2} = \sum_{j=1}^{n-1} \, \left(
\begin{array}{c}
n-2\\
j-1
\end{array} \right)
\, \rho^{j-1} \; {\mathcal P}_j(p) ,
\label{Tnrho}
\end{equation}
with ${\mathcal P}_1(p) = p$, ${\mathcal P}_2(p) = p^2 - p^3$, and ${\mathcal P}_3(p) = 2
\, p^3 - 7 \, p^4 +7 \, p^5 - 2 \, p^6$, etc. We can then invert
eq.~(\ref{Tnrho}) given for $T_n(p,\rho=1)$ to obtain all the ${\mathcal
P}_j(p)$'s,  using the inversion formula
\begin{equation} {\mathcal P}_j(p) =
\sum_{k=1}^{j} \; (-1)^{j+k} \, \left(
\begin{array}{c}
j - 1\\
k - 1
\end{array} \right)
\, T_{k+1}(p,\rho=1) .
\label{polTnrho}
\end{equation}
Combining Eqs.~(\ref{Tnrho}) and (\ref{polTnrho}) gives an easily computable
expression for $T_n(p,\rho)$.


\end{document}